\documentstyle[preprint,aps,eqsecnum,prd,epsf,rotate,floats,times]{revtex}

\begin{document}
\draft

\tighten

\newcommand{\loms}{\lambda_{\rm OMS}}
\newcommand{\hloms}{\hat{\lambda}_{\rm OMS}}
\newcommand{\lams}{\lambda_{\overline{\rm MS}}}
\newcommand{\hlams}{\hat{\lambda}_{\overline{\rm MS}}}
\newcommand{\hlam}{\hat{\lambda} }
\newcommand{\pr}{Phys.\ Rev.\ }
\newcommand{\np}{Nucl.\ Phys.\ }
\newcommand{\phrl}{Phys.\ Rev.\ Lett.\ }
\newcommand{\phl}{Phys.\ Lett.\ }
\newcommand{\cmp}{Comm.\ Math.\ Phys.\ }
\newcommand{\zp}{Z.\ Phys.\ }
\newcommand{\prp}{preprint }

\title{
\vskip-4cm{\baselineskip14pt
\centerline{\normalsize\hskip12.5cm hep-ph/9511407}
\centerline{\normalsize\hskip12.5cm TUM--HEP--224/95}
\centerline{\normalsize\hskip12.5cm TUM--T31--90/95}
\centerline{\normalsize\hskip12.5cm November 1995}}
\vskip2cm
Higgs Sector Renormalization Group in the $\overline{\rm MS}$ and
${\rm OMS}$ Scheme: \\
The Breakdown of Perturbation Theory for a Heavy Higgs}
\author{
Ulrich Nierste\thanks{Electronic address: nierste@physik.tu-muenchen.de}\,
and
Kurt Riesselmann\thanks{Electronic address: kurtr@physik.tu-muenchen.de}\,
}
\address
{Physik-Department T30, Technische Universit\"at M\"unchen,\\
James-Franck-Stra\ss e, 85747 Garching b.\ M\"unchen, Germany}

\maketitle

\begin{abstract}
We discuss different aspects of the Higgs self-interaction in the
$\overline{\rm MS}$ and the on-mass-shell (OMS) scheme. The running
coupling $\lambda(\mu)$ is investigated in great detail. The
three-loop coefficient of the $\beta$-function in the OMS scheme is
derived, and the three-loop running coupling is calculated.
The breakdown of perturbation theory for large Higgs
masses $M_H$ is analyzed in three physical observables for which
two-loop results are known.  Requiring the dependence on the
renormalization scale to diminish order-by-order in $\lambda$, we find
that perturbation theory breaks down for $M_H=$ O(700 GeV) in Higgs
decays. Similarly, $M_H$ must be smaller than O(400 GeV) for
perturbatively calculated cross sections to be trustworthy up to cm
energies of O(2~TeV). If the Higgs sector shall be perturbative up to
the GUT scale, the Higgs must be lighter than O(150 GeV).  For the
two-loop observables examined, the apparent convergence of the
perturbation series is better in the OMS scheme than in the
$\overline{\rm MS}$ scheme.
\end{abstract}

\pacs{PACS numbers: 14.80.Bn, 11.10.Hi, 11.10.Jj, 11.15.Bt, 11.15.Ex}

\newpage

\section{INTRODUCTION}
One of the least tested sectors of the Standard Model of
elementary particle physics is the Higgs sector
generating the masses of all particles via the mechanism
of spontaneous symmetry breaking.
This mechanism implies that the self-coupling $\lambda $ of the
Higgs particle is proportional to the square of its mass, $M_H$. Hence
a heavy Higgs particle may cause the breakdown of perturbation
theory in~$\lambda$.

When calculating decay rates or cross sections beyond the tree level
term of the perturbation series one must specify the {\em
renormalization scheme}\/ to define the coupling constants and the
particle masses appearing in the analytic expressions. In the Higgs
sector one usually adopts the {\em on-mass-shell scheme}\/ (OMS): Here
the squared mass coincides with the physical pole of the propagator,
the vacuum expectation value $v=246$ GeV of the Higgs field is
renormalized such as to cancel tadpole contributions, and the coupling
is chosen to satisfy
\begin{eqnarray}
\loms &=& \frac{M_H^2}{2 v^2} \; = \;
 \frac{G_F \, M_H^2}{\sqrt{2}}  \label{defloms}
\end{eqnarray}
to all orders in perturbation theory.\footnote{We consistently
neglect all gauge coupling corrections.
$\Delta r$ is hence taken to be zero.} Here $G_F$ is the Fermi
constant.  When discussing the perturbation series of physical
observables one must first distinguish processes involving two largely
separated mass scales from those containing only one scale, $M_H$.
Prototypes of the first species are cross sections at LHC
energies.  They contain potentially large logarithmic terms $\lambda
\ln (\sqrt{s}/ M_H) $ with $\sqrt{s}$ being the energy of the process.
These terms may spoil the smallness of radiative corrections. Yet they
can be summed to all orders in perturbation theory with the help of
{\em renormalization group}\/ (RG) methods. The corresponding equation
in the OMS is the {\em Callan-Symanzik equation}, which can be solved
in the limit of large $\sqrt{s}$ when all particle masses can be
neglected.  The second kind of physical observables are one-scale
processes. Examples are two-body decay rates of the Higgs particle
into (almost) massless particles.  We will see in the following that
RG methods are also a useful tool to judge the accuracy of
perturbative results in these cases.

Due to (\ref{defloms}) the use of the OMS appears natural, because
$\loms$ is directly related to physical observables. Yet it is also useful
to consider mass independent renormalization schemes,
the most prominent example being the $\overline{\rm MS}$-scheme.
The reasons are the following:
\begin{itemize}
\item[i)] Mass independent schemes allow for an exact solution of the
          RG equations, i.e.\ mass effects can be systematically
          included.
\item[ii)] The analysis of scheme dependences provides a test of
           the reliability of perturbation theory since the
           results obtained to order $\lambda^n$ in different schemes
           formally differ by terms of order $\lambda^{n+1}$.
\item[iii)] Results obtained in mass independent schemes involve an
           arbitrary parameter, the {\em renormalization scale}\/
           $\mu$. For perturbation theory to work it is necessary that
           the dependence of the result on $\mu$ diminishes order by
           order in perturbation theory. We will use this fact
           extensively in the discussion of the breakdown of
           perturbation theory.
\end{itemize}
The paper is organized as follows: In the following section we
discuss the running coupling constant in the two-loop approximation.
In Sect.~\ref{sect:scheme} we give the two-loop relation between the
quartic Higgs coupling in the OMS and the $\overline{\rm MS}$
scheme. We calculate the three-loop OMS coefficient
of the $\beta$ function as well as the coefficients of the leading and
next-to-leading logarithms. Subsequently, we investigate the scheme
and scale dependence of two-loop physical observables using the
three-loop running coupling. Special attention is paid to the
breakdown of perturbation theory for heavy Higgs masses.
We start by examining the bosonic and
fermionic Higgs decay widths (Sect.~\ref{sect:scale}). Finally, we look at
scattering processes which involve the longitudinally polarized gauge
bosons $W_L^\pm$ and $Z_L$. These scattering processes give the most
stringent bounds on a perturbative Higgs mass (Sect.~\ref{sect:cross}).

\vspace{2cm}

\section{PERTURBATIVE RUNNING COUPLING AT DIFFERENT ORDERS}
\label{sect:rge}

We first discuss the coupling constant $\lambda (\mu)$,
whose running  is encoded in the $\beta$-function.
To three loops, the beta function of the Higgs quartic coupling is defined as
\begin{eqnarray}
\label{beta}
\beta&\equiv&\mu\frac{d\lambda}{d\mu}\equiv\frac{\beta_0}{16\pi^2}\lambda^2
+\frac{\beta_1}{(16\pi^2)^2}\lambda^3
+\frac{\beta_2}{(16\pi^2)^3}\lambda^4
+{\rm O}\biggl(\lambda^5\biggr)\,.
\end{eqnarray}
We neglect all contributions from gauge and Yukawa couplings. This is
an excellent approximation for large values of~$\lambda$. To two loops
the coefficients are~\cite{jack}
\begin{eqnarray}
&&\beta_0=24\,,\qquad\beta_1=-13\beta_0 = -312    \,.
\end{eqnarray}
The three-loop coefficient $\beta_2$ is scheme dependent. We restrict
the discussion of this section to the scheme independent two-loop
results and
return to $\beta_2$ in Sect.~\ref{sect:scheme}.

Equation (\ref{beta}) is valid in any mass independent scheme with
$\mu$ being the renormalization scale accompanying the coupling
constant in the Lagrangian. In the OMS scheme, the Callan-Symanzik
equation describes the response of some Green's function to the
scaling of its external momenta according to $p_i \rightarrow
\mu/\mu_0 \cdot p_i$ which is related to the corresponding scaling of
the coupling given by Eq.~(\ref{beta}).  The values of $\beta_0$ and
$\beta_1$ are scheme independent, so that we don't choose a specific
renormalization scheme until later.  The determination of $\lambda
(\mu)$ proceeds in two steps: First at some initial scale $\mu_0
\approx M_H$ the coupling $\lambda(\mu_0)$ is obtained from
(\ref{defloms}) or an equivalent relation, if the scheme under
consideration is not the OMS scheme.\footnote{For the $\overline{\rm
MS}$ scheme, the expression for $\lambda(\mu_0)$ is derived in
section~\ref{sect:scheme}, see Eq.~(\ref{coup}).}  Then (\ref{beta})
is solved for $\lambda(\mu)\equiv
\lambda [ \lambda(\mu_0), \mu/\mu_0 ]$.  The question whether perturbation
theory works in a specific physical process therefore depends on the two
parameters $\lambda (\mu_0)$ and $\mu/\mu_0$,
which are related to the physical parameters $M_H$ and the energy
$\sqrt{s}$ of the process. The cause of a possible breakdown of perturbation
theory is twofold:
First, the larger the Higgs mass, $M_H$, the larger is
$\lambda (\mu_0)$ due to (\ref{defloms}). Second, the resummation of
possibly large logarithms, $\ln(\sqrt{s}/M_H)$, introduces
the running coupling $\lambda (\mu) $ with $\mu \approx
\sqrt{s}$, and $\lambda (\mu) $
increases for increasing $\mu\approx \sqrt{s}$.

Defining $\lambda$ at some initial scale $\mu_0$, the
solution of (\ref{beta}) yields $\lambda(\mu)$ at any other scale
$\mu$. The expansion of $\lambda ( \mu )$ around $\lambda ( \mu_0 )$
shows that $\lambda(\mu)$ resums powers of the logarithm $\ln
(\mu/\mu_0)$ multiplied by powers of $\lambda( \mu_0 )$. The
coefficients of the leading logarithms (LL), $\lambda^{n+1} (\mu_0)
\ln^n ( \mu/\mu_0) $, depend only on $\beta_0$, those of the
next-to-leading logarithms (NLL) depend on both $\beta_1$ and
$\beta_0$, and so on. We will give these coefficients explicitly
later. The first point we want to stress is that beyond the one-loop
approximation there are several different solutions of
Eq.~(\ref{beta}) due to the truncation of the perturbation series. All
of them agree to the order at which  the perturbative series of the
beta function, (\ref{beta}),
is truncated, but they differ by terms of the neglected order.
For a viable perturbative treatment, these higher-order differences
should be negligible. We now look at four different solutions for the
running coupling $\lambda(\mu)$.

Setting simply the coefficients of the neglected terms in (\ref{beta})
equal to zero and integrating the remaining expressing exactly, we
obtain an implicit equation for $\lambda(\mu)$:
\begin{eqnarray}
\frac{\lambda(\mu_0)}{\lambda(\mu)}&=& 1 -
  \beta_0 \hat{\lambda} (\mu_0) \ln \left( \frac{\mu}{\mu_0}  \right)
+ \frac{\beta_1}{\beta_0}\hat{\lambda}(\mu_0)
\ln\Biggl(
   \frac{  ( \beta_0 + \beta_1 \hlam (\mu ) )  \hlam (\mu_0)}{
           ( \beta_0 + \beta_1 \hlam (\mu_0) ) \hlam (\mu ) }
   \Biggr)
\label{analint}
\end{eqnarray}

in the NLL approximation, and
\begin{eqnarray}
\frac{\lambda(\mu_0)}{\lambda(\mu)}&=& 1 -
  \beta_0 \hat{\lambda} (\mu_0) \ln \left( \frac{\mu}{\mu_0}  \right)
+ \frac{\beta_1}{\beta_0}\hat{\lambda}(\mu_0)
\ln\Biggl(
   \frac{ \hlam (\mu_0) }{\hlam (\mu )}
   \Biggr)
\nonumber \\
&&  + \frac{\beta_0 \beta_2^2\hlam (\mu_0)}{ \lambda_b -\lambda_a }
\left[  \frac{1}{\lambda_b^2}
   \ln \frac{\lambda_b - \hlam (\mu) }{\lambda_b - \hlam (\mu_0) }
  - \frac{1}{\lambda_a^2}
     \ln \frac{\lambda_a-\hlam (\mu) }{\lambda_a - \hlam (\mu_0) }
\right]
   \label{analint2}
\end{eqnarray}
with
\begin{eqnarray}
\lambda_{a/b} &=& - \frac{\beta_1}{2 } \pm \frac{1}{2 }
  		\sqrt{\beta_1^2 - 4 \beta_0 \beta_2} \nonumber
\end{eqnarray}
in the next-to-next-to-leading log (NNLL)
approximation.\footnote{Setting $\beta_2=0$, Eq.~(\ref{analint2})
reduces to (\ref{analint}), of course.}
Here (\ref{analint}) is the two-loop result published
previously~\cite{lend}. In (\ref{analint}) and (\ref{analint2})
the useful abbrevation $\hlam = \lambda/(16 \pi^2)$ has been used.  To
obtain $\lambda( \mu) $ one has to solve (\ref{analint}) or
(\ref{analint2}) by numerical methods. Direct numerical integration of the
original equation, (\ref{beta}), yields the same result for
$\lambda(\mu)$.  Let us call this form the {\em
naive}\/ solution. It would be the exact result if the neglected
coefficients of the $\beta$-function were really identical to
zero. Yet (\ref{analint}) and (\ref{analint2}) contain logarithmic
terms belonging to the neglected higher orders of RG improved
perturbation theory. E.g., the naive NLL result for $\lambda (\mu)$,
(\ref{analint}), partially resums NNLL logarithms $\lambda^{n+3}
(\mu_0) \ln^n ( \mu/\mu_0) $. However, there are further NNLL terms
from irreducible three-loop contributions. Therefore, the naive NLL
result contains an inconsistent resummation of NNLL terms.  Similarly,
Eq.~(\ref{analint2}) includes an inconsistent resummation of
higher-order terms, even though it is correct to the order
considered.  This fact is true whenever beta functions are integrated
numerically at and beyond two loops.

For a {\it consistent} result, one needs to ignore terms of the
neglected order in $\lambda$ when integrating (\ref{beta}). This
yields the consistent NNLL result
\begin{eqnarray}
\frac{\lambda(\mu_0)}{\lambda (\mu)}&=&
1 -\beta_0\hat{\lambda}(\mu_0)\ln(\frac{\mu}{\mu_0})
+ \frac{\beta_1}{\beta_0}\hat{\lambda}(\mu_0)
\ln\biggl( \frac{\lambda(\mu_0)}{\lambda (\mu)} \biggr)
+ \underline{\rule[-17pt]{0pt}{40pt}
              \frac{\beta_1^2- \beta_0 \beta_2 }{\beta_0^2}
              \hat{\lambda}(\mu_0)
              \left[ \hlam (\mu) - \hlam (\mu_0 )  \right] }\,.
\label{consist2lp}
\end{eqnarray}
The consistent NLL result is obtained from (\ref{consist2lp}) by
dropping the underlined terms. The final result for the consistent
solution is obtained by solving Eq.~(\ref{consist2lp}) numerically.

Third we solve (\ref{consist2lp}) iteratively by first substituting
the one-loop result for $\lambda(\mu)$ into the RHS of the equation
above and then repeating this step with the result of the first substitution.
This yields the {\it iterative} answer
\begin{eqnarray}
 \lambda (\mu) &=&
\lambda(\mu_0) \left\{  \rule[-17pt]{0pt}{40pt}
 1 -\beta_0 \hlam ( \mu_0 ) \ln \left( \frac{\mu}{\mu_0} \right)
\right.
\nonumber \\
&&  + \frac{\beta_1}{\beta_0}\hlam ( \mu_0 )
  \;\ln\! \left[\; 1 -\beta_0 \hlam (\mu_0) \ln \frac{\mu}{\mu_0}
   \underline{\rule[-17pt]{0pt}{40pt}  + \frac{\beta_1}{\beta_0} \hlam (\mu_0)
    \ln \left( 1- \beta_0 \hlam (\mu_0) \ln \frac{\mu}{\mu_0}   \right)   }\;
\right]
\nonumber \\
&& \left.  \underline{ \rule[-17pt]{0pt}{40pt}
+ \frac{\beta_1^2-\beta_0 \beta_2}{\beta_0 } \:
     \hlam ^3 (\mu_0)  \ln \frac{\mu }{\mu_0} \cdot
      \left( 1- \beta_0 \hlam(\mu_0) \:\ln \frac{\mu}{\mu_0}  \right)^{-1}
   }  \right\} ^{-1} . \label{pert2lp}
\end{eqnarray}
We stress that no further expansions in $\hlam (\mu_0)$ are possible,
because each $\hlam(\mu_0)$ multiplies a large logarithm
$\ln(\mu/\mu_0)$.  Eq.~(\ref{pert2lp}) is the NNLL result, and the NLL
expression is again obtained by dropping the underlined terms.

The fourth solution of (\ref{beta}) is constructed in analogy to QCD:
The integration constant obtained in the integration of (\ref{beta})
can be absorbed into a scale parameter $\Lambda_H$:
\begin{eqnarray}
\frac{\lambda(\mu)}{16\pi^2}=
\frac{2}{\beta_0\ln(\frac{\Lambda_H^2}{\mu^2})}
&&\left[1 -\frac{2 \beta_1}{\beta_0^2}
          \frac{\ln[\ln(\frac{\Lambda_H^2}{\mu^2})]}{
                \ln(\frac{\Lambda_H^2}{\mu^2})}
       \underline{\rule[-21pt]{0pt}{40pt}
       + \frac{4 \beta_1^2}{\beta_0^4}
         \frac{\ln^2 [ \ln \frac{\Lambda_H^2}{\mu^2} ] }{
               \ln^2(\frac{\Lambda_H^2}{\mu^2}) }
              - 4 \frac{\beta_1^2}{\beta_0^4} \frac{\ln [ \ln(
       \frac{\Lambda_H^2}{\mu^2}) ] }{\ln^2
             (\frac{\Lambda_H^2}{\mu^2}) } } \right.\nonumber \\
&&\;\; \left.\underline{\rule[-21pt]{0pt}{40pt}
      + 4 \frac{ \beta_0 \beta_2 - \beta_1^2}{\beta_0^4}
          \frac{1}{\ln^2(\frac{\Lambda_H^2}{\mu^2}) } }
+{\rm O}\biggl(\frac{\ln^3[\ln( \frac{\Lambda_H^2}{\mu^2} )]}{
              \ln^3(\frac{\Lambda_H^2}{\mu^2})}\biggr)
\right].
\label{lamdef}
\end{eqnarray}
We call this result the {\em QCD-like}\/ solution.  The above equation
defines the scale parameter $\Lambda_H$, and it is written such that
no term of the form $\mbox{\it const.}/\ln (\Lambda_H^2/\mu^2) $
appears in the square brackets. This is identical to the definition of
the QCD scale parameter $\Lambda^{\overline{\rm MS}}_{\rm QCD}$
\cite{bbdm}.  The definition of the NLL parameter $\Lambda_H^{\rm
NLL}$ is obtained by
dropping the underlined terms.  The above definition of $\Lambda_H$
works in any renormalization scheme. The actual numerical evaluation
of $\Lambda_H$ depends on the boundary value $\lambda (\mu_0)$, which
is scheme dependent.  If for example $\lambda(\mu)$ on the LHS of
(\ref{lamdef}) is given in the $\overline{\rm
MS}$-scheme, $\Lambda_H$ on the RHS equals $\Lambda_H^{\overline{\rm
MS}}$.

There are two important
differences between the Higgs sector and QCD: First, Eq.~(\ref{lamdef})
holds for $\mu \ll \Lambda_H $, while in an asymptotically free theory
like QCD the analogue of (\ref{lamdef}) is valid for $\mu \gg
\Lambda_{\rm QCD} $.
Second, $\Lambda_{\rm QCD}$ is the only fundamental scale parameter of
QCD (with massless quarks), while in the Higgs sector $\Lambda_H$ is
related to the Higgs mass (see Table~\ref{lamtable}).  Yet if the
Higgs mass turns out to be large, one will have to parametrize
non-perturbative effects in terms of an effective Lagrangian, and
$\Lambda_H$ will be the natural scale entering the effective
couplings.
In QCD perturbation theory breaks down
for $\mu \lesssim 3\,\Lambda^{\overline{\rm MS}}_{\rm QCD} $ and binding
energies equal a few times $\Lambda_{\rm QCD}$. In the Higgs sector
the breakdown of perturbation theory likewise occurs for $ \mu \gtrsim
\Lambda_H/3$.  In Table~\ref{lamtable} we tabulate the relation
between $M_H$ and $\Lambda_H$. When the OMS scheme is adopted, $M_H$
in Table~\ref{lamtable} is the physical mass (see (\ref{defloms})).
In other schemes the tabulated values for $M_H$ correspond to the
tree-level relation $\lambda (\mu_0 =M_H^{\mbox{\scriptsize tree }})=
(M_H^{ \mbox{\scriptsize tree } })^2/(2 v^2)$ and --- in contrast to
the OMS scheme --- radiative corrections to this relation have to be
taken into account in order to obtain the physical Higgs mass from
$M_H^{ \mbox{\scriptsize tree } }$. For the $\overline{\rm MS}$-scheme
this relation is encoded in (\ref{coup}) of Sect.~\ref{sect:scheme}.

\begin{table}[tbh]
\begin{displaymath}
  \begin{array}{|c|c|c|c|c|c|}
  \hline
   M_H^{\protect\mbox{\protect\scriptsize tree} } (\text{GeV})  &
   \Lambda_H^{\rm LL}  (\text{GeV})  &
   \Lambda_H^{\rm NLL}  (\text{GeV}) &
   \Lambda_{H, {\rm OMS}}^{\rm NNLL}  (\text{GeV}) &
   \Lambda_{H, \overline{\rm MS}}^{\rm NNLL}  (\text{GeV}) &
   \mu^{\rm NLL}_{\rm max}  (\text{GeV})
  \\ \hline
100. & 4 \cdot 10^{36} & 6 \cdot 10^{37} & 7 \cdot 10^{37} &
7 \cdot 10^{37}& 2.2 \cdot 10^{36}  \\
200. & 9 \cdot 10^{10} & 6 \cdot 10^{11} & 7 \cdot 10^{11} &
7 \cdot 10^{11} & 5.1 \cdot 10^{10} \\
300. & 2.1 \cdot 10^6 & 8.7 \cdot 10^6 & 1.0 \cdot 10^7 & 1.0
\cdot 10^7& 1.2 \cdot 10^6 \\
400. & 5.8 \cdot 10^4 & 1.8 \cdot 10^5 & 2.0 \cdot 10^5 & 2.1
\cdot 10^5& 3.4 \cdot 10^4  \\
500. & 1.2 \cdot 10^4 & 2.9 \cdot 10^4 & 3.3 \cdot 10^4 & 3.6
\cdot 10^4& 7.0 \cdot 10^3  \\
600. & 5.51 \cdot 10^3 & 1.13 \cdot 10^4 & 1.25 \cdot 10^4 & 1.40
\cdot 10^4 & 3.2 \cdot 10^3 \\
700. & 3.57 \cdot 10^3 & 6.45 \cdot 10^3 & 6.92 \cdot 10^3 & 7.99
\cdot 10^3& 2.1 \cdot 10^3 \\
800. & 2.78 \cdot 10^3 & 4.52 \cdot 10^3 & 4.71 \cdot 10^3 & 5.64
\cdot 10^3& 1.6 \cdot 10^3\\
900. & 2.41 \cdot 10^3 & 3.57 \cdot 10^3 & 3.60 \cdot 10^3 & 4.52
\cdot 10^3 & 1.40 \cdot 10^3 \\
1000. & 2.22 \cdot 10^3 & 3.03 \cdot 10^3 & 2.94 \cdot 10^3 & 3.91
\cdot 10^3&  1.29 \cdot 10^3\\
1100. & 2.13 \cdot 10^3 & 2.68 \cdot 10^3 & 2.48 \cdot 10^3 & 3.57
\cdot 10^3& 1.24 \cdot 10^3 \\
1200. & 2.09 \cdot 10^3 & 2.41 \cdot 10^3 & 2.18 \cdot 10^3 & 3.37
\cdot 10^3& 1.21 \cdot 10^3 \\
  \hline
  \end{array}
\end{displaymath}
\caption{The values for the scale parameter $\Lambda_H$
obtained from (\protect\ref{lamdef}) for
$\mu_0=M_H^{\protect\mbox{\protect\scriptsize tree} }$.
$M_H^{\protect\mbox{\protect\scriptsize tree} }$ in the left column
corresponds to $M_H^{\protect\mbox{\protect\scriptsize tree} }= v
\protect\sqrt{2
\lambda(\mu_0=M_H^{\protect\mbox{\protect\scriptsize tree} })} $.
For $\lambda=\protect\loms$ this relation receives no radiative
corrections. If the coupling is defined in the $\overline{\rm
MS}$-scheme, use (\protect\ref{coupnum}) first to calculate $\protect
\lams (\mu_0= M_H^{\protect\mbox{\protect\scriptsize tree} })$
from the physical Higgs mass to the desired order. Then use
the previous equation  to obtain $M_H^{\protect\mbox{\protect\scriptsize
tree} }$.  $\Lambda_H^{\rm LL}$ corresponds to the location of the
one-loop Landau pole.  The difference between the fourth and fifth
column is caused by the scheme dependence of $\beta_2$.  The last
column contains the two-loop values of $\mu$ for which the iterative
solution, (\protect\ref{pert2lp}), assumes its maximum.  }
\label{lamtable}
\end{table}

We now completed the definition of the four different solutions for
the running Higgs coupling: naive, consistent, iterative, and QCD-like
solution.  It is interesting to note that at one loop all four
solutions are identical.\footnote{To see this, set $\beta_1=\beta_2=0$
in the four different solutions.}  This is clearly not the case
anymore at two-loop and higher orders, and we will discuss the
differences below. Let us stress again that {\it all} solutions
correctly sum the large logarithm $\ln (\mu/\mu_0)$ within the
calculated order.  The difference between these solutions is of the
neglected order as can be seen when expanding the solutions in
$\lambda(\mu_0)$.  If perturbation theory is applicable this
difference should be numerically small, giving a very simple criterion
to find the values of $M_H$ and $\mu$ beyond which perturbation theory
clearly fails.

\begin{figure}[tbh]
\vspace*{13pt}
\centerline{ \epsfysize=4.8in
  \rotate[l]{\epsffile{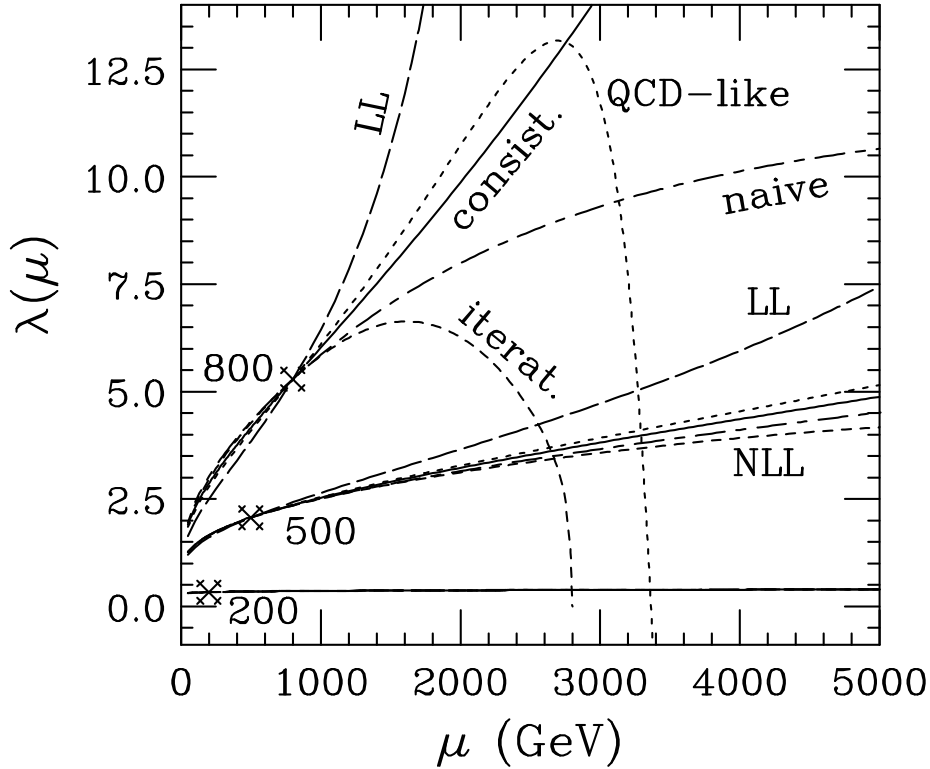}} }
\vspace{0.15in}
\caption{
Different solutions of the one-loop and two-loop RGE equations for the
Higgs quartic coupling $\lambda$: the one-loop solution (long dashes),
the two-loop naive solution according to (\protect\ref{analint})
(long-short dashes), the two-loop consistent solution of
(\protect\ref{consist2lp}) (solid line), the two-loop iterative
solution (\protect\ref{pert2lp}) (short dashes), and the two-loop
QCD-like solution expanded in powers of $1/\ln(\Lambda_H^2/\mu^2)$
given in (\protect\ref{lamdef}) (dots).  The crosses show the
normalization point:
$\lambda(\mu=M_H^{\protect\mbox{\protect\scriptsize tree}
})=\lambda_0$. To allow for a meaningful comparison of the different
orders in perturbation theory, we choose $\lambda_0$ to be the same at
one and two loops. The curves are labeled by the tree-level Higgs mass
(in GeV), corresponding to the tree level relation $\lambda_0= \left(
M_H^{\protect\mbox{\protect\scriptsize tree} }\right)^2/(2v^2)$.  In
renormalization schemes different from the OMS, radiative corrections
to this relation must be included (see text).  } \label{rgecurves}
\end{figure}

In Fig.~\ref{rgecurves}, we compare the $\mu$ dependence of the four
solutions at one and two loops.  We choose the three mass values
$M_H=200$, $500$, and $800$ GeV.  To obtain a meaningful comparison
between the one- and two-loop results for a given value of $M_H$, we
take for both orders the same expansion parameter $\lambda_0\equiv\lambda
(\mu_0) $, and choose it to be defined by the tree-level relationship
$\lambda_0=\left(M_H^{\rm tree}\right)^2/(2v^2)$.  This has the
additional advantage that we
do not yet have to specify a renormalization scheme in
Fig.~\ref{rgecurves}.  Since the first two
coefficients of the $\beta$-function are scheme independent, the whole
scheme dependence resides in the relation between
the tree-level mass (which labels the different curves in
Fig.~\ref{rgecurves}) and the physical Higgs mass.  In the OMS, the
two mass definitions coincide.

At one loop, the four solutions are identical (long dashed
curve). The coupling $\lambda^{\rm LL}(\mu)$ approaches infinity as
$\mu\rightarrow\Lambda^{\rm LL}_H$, which is usually referred to as
the Landau pole~\cite{landau}.
At two loops the breakdown of perturbation theory is clearly visible
in the two-loop curves for $M_H=800$ GeV.
It manifests itself in a very different behaviour of the four solutions,
although they all are defined to have the same value for $\mu_0=M_H$.
The one-loop Landau pole has completely vanished from all two-loop solutions.
The  naive (long-short dashes in Fig.~\ref{rgecurves}) and consistent
(solid) solutions always have positive
slopes in accordance with the positiveness of the perturbative
$\beta$-function, Eq.~(\ref{beta}).
The consistent
solution can be looked at as the ``average'' of the one-loop solution
and the two-loop naive solution.   The naive solution
approaches the fix-point value $\lambda^{\rm NLL}_{\rm max}=12.147$,
the zero of the two-loop beta function.  This simply reflects the fact
that one cannot use perturbation theory to gain information about
large values of the coupling. This is even more evident from the other
two solutions. The QCD-like result for $\lambda^{\rm
NLL}(\mu)$ has a global maximum.
Interestingly enough, for values of $M_H>1263$ GeV, the value
$\lambda_0=M_H^2/(2v^2)$ is larger than the maximal possible value of the
QCD-like running coupling, so that $\lambda(\mu=M_H)=\lambda_0$ can no
longer be satisfied. The
unphysical maximum of the QCD-like solution
appears for $\mu^{\it QCD}_{\rm max}=\Lambda^{\rm NLL}_H/1.7$, and
perturbation breaks down well before  $\mu$ is equal to $\Lambda_H$.
The iterative solution features a maximum at
\begin{equation}
\label{maxiter}
\mu_{\rm max}^{iter} = M_H \exp\left(
           \frac{1+\frac{\beta_1}{\beta_0}\hat\lambda_0}
{\beta_0\hat\lambda_0}\right)
= 0.58 M_H \exp\left(\frac{6.58}{\lambda(\mu_0)}\right)
\, ,
\end{equation}
which is located at  $1615$ $(7033)$ GeV for $M_H=800$ $(500)$ GeV.
The two-loop iterative solution approaches zero at the same value of
$\mu$ at which the one-loop solution has its Landau pole.

The fact that some perturbative solutions of the two-loop running coupling show
a maximum gives a first criterion
for the breakdown of the two-loop perturbative treatment.
{}From the definition of the
$\beta$ function, Eq.~(\ref{beta}), we know that in the perturbative regime
the slope of the running coupling has to be positive as the one-loop term
dominates over the two-loop term. If the iterative solution for
$\lambda(\mu)$ has negative slope, then perturbation theory is not valid
anymore.
Hence Eq.~(\ref{maxiter}) is possibly a measure of the range of
$\mu$ for which perturbation theory is meaningful.

The previous discussion of the four solutions is valid for any value
of $M_H$. Yet if we restrict ourselves to values $\mu< 5000$ GeV,
features like poles and maxima are only apparent for $M_H=800$ GeV,
and not for $500$ and $200$ GeV (see Fig.~\ref{rgecurves}).  For the
lower mass values one has to go to (much) larger values of $\mu$  to observe
the breakdown of perturbation theory.  In fact, if $M_H=200$ GeV all
four solutions of the running couping agree extremely well for
$\mu<5000$ GeV. Choosing $M_H=500$, however, already results in an
uncertainty between the different two-loop solutions of about 25\% at
$\mu= 5$ TeV, and the one-loop solution is more than 50\%
larger than the two-loop solution at $\mu=5$ TeV.

If the Standard Model Higgs sector is to remain perturbative up to much
higher energy scales the restrictions on $M_H$ get much more
severe~\cite{petron}.  To avoid the Landau pole and keeping the
coupling $\lambda$ perturbative, one can define an embedding scale
below the Landau pole and require the running coupling to be smaller
than a certain value. For example, taking the embedding scale to be
$10^{10}$ GeV the authors of \cite{lindner}
implement the ``automatic fixing procedure''
requiring (in our notation) $\lambda(10^{10}{\rm
GeV})<\pi^2$ and find an upper bound of $M_H\approx 230$ GeV when
neglecting the top-quark Yukawa coupling.
Using Eq.~(\ref{maxiter}) and requiring
$\mu_{\rm max}^{iter}$ to be equal or larger than $10^{10}$ GeV, we
find the upper limit $M_H\approx 210$ GeV.

The analysis of the running coupling alone can clearly only yield a
necessary criterion for the validity of perturbation theory. Taking
$M_H=800$ GeV, we can be sure that perturbation theory breaks down for
any cross section with $\sqrt{s} \approx 2$ TeV, because this energy is
too close \cite{luewei} to the scale parameter $\Lambda_H$ and
already larger than the corresponding maximum of $\mu$, $\mu_{\rm
max}^{iter}(M_H=800 {\rm GeV})\approx 1600$ GeV.
For $M_H\leq500$ GeV, the perturbative RG treatment of the running
coupling could be trustworthy up to values of  $\mu_{\rm
max}^{iter}\approx 7$ TeV, sufficient
for studying Higgs physics at the LHC.  Nevertheless, choosing
$M_H=500$ GeV the running coupling
$\lambda^{\rm NLL}(\mu\!=\!7\,\mbox{TeV})=4.4$ is sizable.

To make a final judgement on whether perturbation theory works for
$M_H=500$ GeV and TeV-energies, one must in addition investigate the
perturbation series of the physical process of interest.
As we will see later on,
the perturbative solution of the running
coupling may show a reasonable convergence, but  the numerical value
of the running coupling is already too large to calculate physical quantities
like cross sections and decay widths in perturbation theory.
This will be the subject of sections~\ref{sect:scale}
and~\ref{sect:cross}.

\section{SCHEME DEPENDENCE: OMS vs. $\overline{\rm \bf MS}$
FORMULATION}
\label{sect:scheme}
So far the discussion of the running coupling has been independent of
a special renormalization scheme. We have already listed the reasons
for studying different renormalization schemes  in the introduction.

In the following we will look at the $\overline{\rm MS}$ and OMS
scheme, and examine the scheme dependence of the coupling and the
three-loop coefficient $\beta_2$ in (\ref{beta}). Before going into
detail we would like to remark two points: First, the OMS coupling
$\lambda_{\rm OMS}$ is to all orders in $\lambda$
directly related to measurable quantities (the muon
lifetime and the Higgs mass) via (\ref{defloms}). This is not so for any other
renormalization scheme, where the RHS of (\ref{defloms}) receives
radiative corrections, which depend on an additional parameter, the
{\em renormalization point}\/ $\mu_0$.  Second the scheme dependence
of $\beta_2$ makes the coupling run differently in different schemes.
One can in general adjust the scheme  such as to achieve any desired
values for $\beta_n$ with $n \geq 2$. A criterion for a ``good
scheme'', however, cannot be founded on the smallness of the running
coupling   alone. Instead one has to consider physical observables,
in which the coefficients of the perturbation series depend on the
scheme as well. This will be done in the following sections.

The starting point of the analysis is the two-loop relation between
the bare and the renormalized coupling:
\begin{eqnarray}
\mu^{-2 \epsilon}\hlam^{\rm bare} =
\hlam &+& \,\hlam^2 \xi^{\epsilon} \,
 \left[\; \frac{\beta_0}{2}   \frac{1}{\epsilon} + c_1 +
 \epsilon c_{11} + \epsilon^2 c_{12} + O( \epsilon^3)  \right]  \nonumber \\
&+& \hlam^3 \xi^{2 \epsilon} \left[
     \;\frac{\beta_0^2}{4} \frac{1}{\epsilon^2} +
     \left( \beta_0 c_1 + \frac{\beta_1}{4}   \right)
     \frac{1}{\epsilon}  + c_2  +\epsilon c_{21} + O( \epsilon^2 )\right]
\nonumber\\
&+& \hlam^4 \xi^{3 \epsilon} \left[
     \;\frac{\beta_0^3}{8} \frac{1}{\epsilon^3} +
 \left(\frac{3}{4} \beta_0^2 c_1 + \frac{7}{24}\beta_0\beta_1
\right) \frac{1}{\epsilon^2}   \right.
\nonumber\\
&&\left.
   \phantom{\hlam^4 }
   \;+
   \left(-\frac{5}{12} \beta_0^2 c_{11}
	+\frac{1}{3} \beta_0 c_{1}^2
	+\frac{7}{6} \beta_0 c_{2}
	+ \frac{7}{12}\beta_1 c_1
	+ \frac{1}{6}\beta_2
   \right)
     \frac{1}{\epsilon}
+ c_3  + O( \epsilon )\right] \,.
\label{z}
\end{eqnarray}
In the $\overline{\rm MS}$-scheme the various quantities are given as
\begin{eqnarray}
&&  \hlam =  \frac{ \lambda_{\overline{\rm MS}}(\mu ) }{16 \pi^2},  \quad \quad
\xi = 4 \pi e^{-\gamma_E}, \quad \quad
c_1=c_2=c_3=c_{11}=c_{12}=c_{21}=0. \nonumber
\end{eqnarray}
In the OMS scheme, they  instead read \cite{maher,gh}:
\begin{eqnarray}
&&  \hlam =  \frac{ \lambda_{\rm OMS} }{16 \pi^2},  \quad
\xi = \frac{4 \pi \mu^2 e^{-\gamma_E}}{M_H^2}, \quad
c_1= 25-3 \pi \sqrt{3}= 8.676\,, \quad
c_2= 378.5\,, \quad
c_{11}=3.821 \,,
\label{omscoeff}
\end{eqnarray}
and $c_3,\; c_{12}$, and $c_{21}$ are unknown. Knowing the bare
coupling to two loops in both the OMS scheme and the $\overline{\rm
MS}$ scheme, we can express the $\overline{\rm MS}$ coupling in terms
of the OMS coupling, and we can calculate the difference between
$\beta_2^{\rm OMS}$ and $\beta_2^{\overline{\rm MS}}$.

\subsection{The $\overline{\rm MS}$-coupling $\lams$
and the renormalization point $\mu_0$ }\label{sect:mscoup}

Since the OMS coupling is a function of $M_H$ (or $v$) and is fixed
to all orders by (\ref{defloms}), the definition of
$\lambda_{\overline{\rm MS}}$ in terms of $\lambda_{\rm OMS}$ is
equivalent to a relation between $\lambda_{\overline{\rm MS}}$ and
$M_H$.  With $c_i$ and $c_{ij}$ refering to the OMS quantities in
(\ref{omscoeff}) we find:
\begin{eqnarray}
\nonumber
\lambda_{\overline{\rm MS}}(\mu_0)  =
\lambda_{\rm OMS}\Biggl[1\, \Biggr.
&+&\, \left( \frac{\beta_0}{2} \ln \frac{\mu_0^2}{M_H^2}
+ c_{1}\right)
\hat{\lambda}_{\rm OMS}\,\\ \nonumber
&+&\, \left( \left(\frac{\beta_0}{2}\right)^2\ln^2
\frac{\mu_0^2}{M_H^2} +
\left(\frac{\beta_1}{2}+\beta_0c_1\right)\ln
\frac{\mu_0^2}{M_H^2} + c_2 -
\beta_0c_{11}\right)\hat{\lambda}_{\rm OMS}^2\\ \nonumber &+&\,
\Biggl( \left(\frac{\beta_0}{2}\right)^3\ln^3
\frac{\mu_0^2}{M_H^2}
+\frac{\beta_0}{2}\left(\frac{5}{4}\beta_1+\frac{3}{2}\beta_0
c_{1}\right) \ln^2 \frac{\mu_0^2}{M_H^2} \Biggr.\\
\nonumber && \;\;\;+ \Bigl( \left( 3 c_{2} -3\beta_0 c_{11}\right)
\frac{\beta_0}{2} + \beta_1 c_{1} + \frac{1}{2} \beta_2^{\rm OMS}
\Bigr)\ln \frac{\mu_0^2}{M_H^2} \nonumber\\ & & \Biggl.\;
\; +\; c_3 +\left(\frac{1}{4}\beta_0 c_{12} -c_{1}c_{11} -
c_{21}\right)\beta_0 -\frac{3}{4}\beta_1 c_{11}
\Biggr)\hat{\lambda}_{\rm OMS}^3\nonumber\\ &+&\Biggl.\,
{\rm O}\left(\hat{\lambda}_{\rm OMS}^4\right) + {\rm O}(\epsilon)
\Biggr],    \label{coup}    \\
=  {\hat\lambda}_{\rm OMS}\Biggl[1\, \Biggr.
&+&\, \left( 12 \ln \frac{\mu_0^2}{M_H^2}  + 8.676\right)
{\hat\lambda_{\rm OMS}}\, \nonumber \\
&+&\, \left( 144\ln^2 \frac{\mu_0^2}{M_H^2}
+ 52.22\ln \frac{\mu_0^2}{M_H^2}
+ 286.8\right){\hat\lambda_{\rm OMS}^2} \nonumber\\
&+&\Biggl.\,
\left( 1728 \ln^3 \frac{\mu_0^2}{M_H^2}
- 932.0 \ln^2 \frac{\mu_0^2}{M_H^2}  + 9736.8 \ln
\frac{\mu_0^2}{M_H^2} + d_{30} \right)  {\hat \lambda }_{\rm OMS}^3
\nonumber\\
&&
  + {\rm O}(\epsilon)
\Biggr], \label{coupnum}
\end{eqnarray}
which agrees to one loop with the result of Sirlin and
Zucchini~\cite{sirzuc}.  The three-loop constant term $d_{30}$
depends on the yet unknown OMS coefficients $c_{12}$, $c_{21}$, and
$c_3$.

For convenience we give the inverse formula of (\ref{coupnum}) as
well:
\begin{eqnarray}
\lambda_{\rm OMS}
= \lambda_{\overline{\rm MS}}(\mu)\Biggl[1\, \Biggr.  &+&\, \left( -12
\ln \frac{\mu^2}{M_H^2}  - 8.676\right)
{\hat\lambda_{\overline{\rm MS}}}(\mu)\, \nonumber \\ &+&\,
\left(144\ln^2 \frac{\mu^2}{M_H^2}
+ 364.2 \ln \frac{\mu^2}{M_H^2}  -136.3 \right)
{\hat\lambda_{\overline{\rm MS}}^2} (\mu)\nonumber\\ &+&\Biggl.\, {\rm
O}\left({\hat\lambda}_{\overline{\rm MS}}^3\right) + {\rm O}(\epsilon)
\Biggr] \,. \label{coupinv}
\end{eqnarray}
Eq.~(\ref{coup}) defines the expansion
parameter $\lambda(\mu_0)$ of the running coupling when using the
$\overline{\rm MS}$ scheme.  It shows that $\lams$ is completely
determined by specifying $M_H$ (and thereby $\loms$) and a
renormalization point $\mu_0$ at which (\ref{coup}) is imposed. The
scale $\mu_0$ is unspecified.  However, to ensure that the logarithms
$\ln (\mu_0/M_H)$ stay small at all orders, it should be chosen of the
order of $M_H$.

Let us emphasize that throughout the paper $M_H$
denotes the pole mass, even when discussing $\overline{\rm MS}$
renormalization, and the vacuum expectation value $v$ is also chosen
as defined in the OMS as in \cite{sirzuc}.  Expressions involving the
running mass of the $\overline{\rm MS}$-scheme can systematically be
expressed in terms of the pole mass $M_H$.  Despite the use of the
pole mass $M_H$, the $\overline{\rm MS}$ renormalization scheme
maintains all the advantages of mass independent schemes mentioned in
the introduction.
For example  $\lams (\mu_0=M_H)= (800 \, {\rm GeV})^2/(2v^2)$
corresponds to a physical Higgs mass of $M_H=720 $ GeV at one loop, and
$M_H=681$ GeV at two loops.

\begin{figure}[tb]
\vspace*{30pt}
\centerline{
\epsfysize=3.8in \rotate[l]{\epsffile{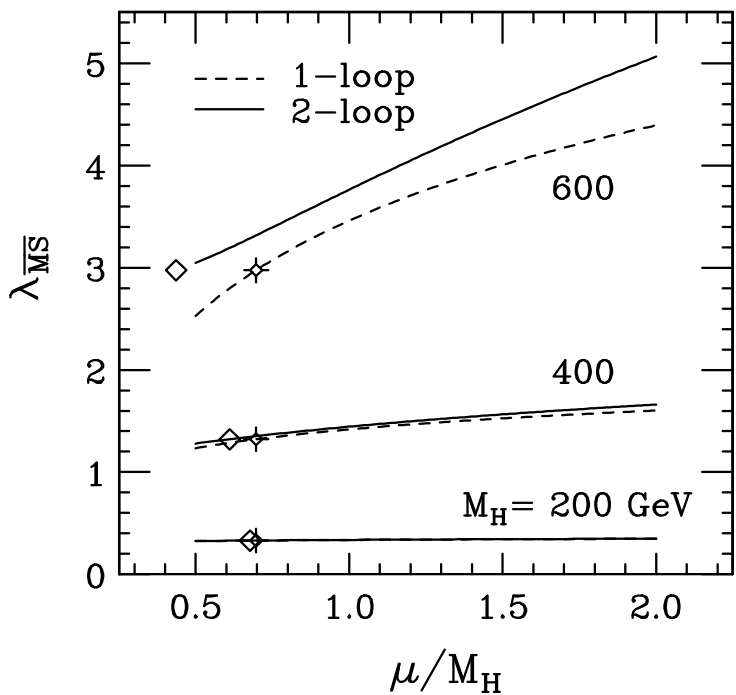}}
}
\vspace{0.15in}
\caption{The ${\overline{\rm MS}}$ coupling, $\lambda_{\overline{\rm MS}}$,
as a function of the
renormalization scale $\mu$, fixing the physical Higgs mass at
$200,\:400$, and $600$ GeV.  The value of $\mu$ at which the two-loop
(one-loop)  ${\overline{\rm MS}}$ coupling
is equal to the $\mu$-independent OMS-coupling, $\lambda_{\rm OMS}$, is
indicated by a diamond (crossed diamond).
}
\label{msbar}
\end{figure}

In Fig.~\ref{msbar} we show the $\overline{\rm MS}$ coupling as a
function of $\mu_0$ for different values of $M_H$, varying $\mu_0$ in
a typical range, $M_H/2 \leq \mu_0 \leq 2M_H$.  We find that the
$\overline{\rm MS}$ coupling is larger than the OMS coupling for most
of the range of $\mu_0$ examined.  If $M_H$ is larger than $400$ GeV,
the value of $\mu_0$ at which $\lambda_{\overline{\rm MS}}(\mu_0)$
equals $\lambda_{\rm OMS}$ changes significantly when going from one
loop to two loops.  At one loop, the two couplings are equal if
$\mu_0\approx 0.7\,M_H$ for any value of $M_H$, because $\beta_0/2
\cdot \ln (0.7^2) = -c_1$.  At two loops, the relation gets a more complicated
mass dependence. We find that for $M_H=400\;(600)$ GeV the two-loop
couplings of the two schemes are equal if
$\mu_0=0.6\,M_H\;(0.4\,M_H)$.

At first sight it appears reasonable to choose $\mu_0$ in such a way
that calculated radiative corrections become small or even zero.  This
criterion of {\em (fastest) apparent convergence}\/ (FAC) has had some
supporters in QCD RG analyses over a decade ago (for a criticism see
\cite{steve}). Also in \cite{sirzuc} the one-loop FAC scale
$\mu_0=0.7 M_H$, at which the one-loop correction in (\ref{coup})
vanishes, has been used to define $\lams$. Yet in the following
we will explain why this choice of the scale is not very useful, in
particular in conjunction with the analysis of the breakdown of
perturbation theory. We can already anticipate this fact from the
above observation that for large values of $M_H$ the two-loop FAC
scale differs  sizably from its one-loop value. For a more
detailed investigation consider  the generalization of  the square bracket
in (\ref{coup}) to the  order $\hloms^N$:
\begin{eqnarray}
\hat{\lambda}_{\overline{\rm MS}}(\mu_0)  =
\hat{\lambda}_{\rm OMS} \left[ 1\, +
    \sum_{n=1}^N \hloms^n \,
     \sum_{k=0}^n d_{nk} \ln ^k \frac{\mu_0^2}{M_H^2} \right] ,
\label{ordern}
\end{eqnarray}
i.e.\ one has $d_{10}=c_1,d_{11}=\beta_0/2 $ and so on.
Application of the renormalization group equation (\ref{beta}) (with
$\mu=\mu_0$) yields recursion relations for the
$d_{nk}$'s. The solution for the coefficients $d_{nn}$ of the leading
logarithms is well-known:
\begin{eqnarray}
d_{nn} &=& \left( \frac{\beta_0}{2}  \right)^n .   \label{llog}
\end{eqnarray}
Similarly  we derive the general relationship
\begin{eqnarray}
d_{n,n-1} &=& c_1 \, n \, \left( \frac{\beta_0}{2}\right)^{n-1}  +
	a_n \, \frac{\beta_1}{2}\, \left( \frac{\beta_0}{2}\right)^{n-2}
    \label{nllog}
\end{eqnarray}
for the next-to-leading logarithms with
\begin{eqnarray}
a_n &=& \sum_{k=1}^{n-1}  \sum_{j=1}^{k} \frac{1}{j} \; = \;
       n \left[ \psi (n+1) + \gamma_E -1   \right] , \nonumber
\end{eqnarray}
where $\psi (x) = \frac{d}{d\,x} \ln \Gamma(x) $.
For $n \leq 3$ the coefficients $d_{nn}$ and $d_{n,n-1}$ have already
been given in (\ref{coup}), where also the NNLL coefficient $d_{31}$
involving $\beta_2$ is displayed.

We remark here that the coefficients (\ref{llog}) and (\ref{nllog})
have the same form for any expansion of a running coupling
$\lambda(\mu)$ in terms of some $\mu$-independent coupling with only
the coefficient $c_1$ changed correspondingly.  For example, the
expansion of $\lams (\mu)$ in terms of $\lams (M_H)$ yields a series
of the form (\ref{ordern}) with $\hloms \rightarrow \hlams (M_H)$ and
$d_{10}=c_1\rightarrow 0$ in (\ref{nllog}).

Now the choice $\mu_0 = M_H$ nullifies all logarithms to all orders in
(\ref{ordern}). The FAC scale $\mu_0=0.7\,M_H$ instead nullifies the
$\hloms^1$-term in (\ref{ordern}) and (\ref{coup}), but the price to
be paid is the appearance of logarithms in higher orders.
Since we know the NLL coefficients $c_2$ and $c_{11}$, we can check
the effect of the FAC scale setting on the $\hloms^2$-term in
(\ref{coup}): For $\mu_0=M_H$ the coefficient of $\hloms^2$ equals
287, but for $\mu_0=0.7\,M_H$ one finds the larger coefficient
324 which increases the impact of higher order terms, especially if
the coupling is large.
This explains the above finding that the FAC scale
changes significantly for large values of $M_H$ when passing from the
one-loop to the two-loop order.  Hence the FAC scale setting pushes
large terms from the calculated orders into the uncalculated higher
orders of the perturbation series. Yet a clever choice of $\mu_0$
should yield the opposite and keep these higher order terms small.

Of course one is not forced to use $\mu_0=M_H$ exactly. But what is
the allowed range for $\mu_0$? Clearly $\ln ( \mu_0^2/M_H^2) $ in
(\ref{ordern}) must be kept small, so that the logarithms do not
become dominant in higher orders. A reasonable interval for $\mu_0$
should therefore obey
\begin{eqnarray}
\left| d_{n,m+1} \ln^{m+1} \frac{\mu_0^2}{M_H^2} \right|
&\lesssim&  \left| d_{nm} \ln^m \frac{\mu_0^2}{M_H^2} \right|    \nonumber
\end{eqnarray}
for $m=0,\ldots n-1$.  Especially the size of the logarithms
should not exceed the constant term $|d_{n0}|$.
With the two-loop information (\ref{coupnum}) at hand we
first compare the size of both $d_{22} \ln^2 (\mu_0^2/M_H^2) $ and
$d_{21} \ln (\mu_0^2/M_H^2) $ with the non-logarithmic term,
$d_{20}= 286.8 $, and find that for $0.5 \leq
\mu_0/M_H \leq 2.0$ their magnitudes do not
exceed $|d_{20}|$. Second, using (\ref{llog})
and (\ref{nllog}) we can compare the leading logarithm $d_{nn} \ln^n
(\mu_0^2/M_H^2)$ with the next-to-leading logarithms $d_{n,n-1} \ln^{n-1}
(\mu_0^2/M_H^2)$ to all orders in perturbation theory obtaining the
smaller range $0.8 \leq \mu_0/M_H \leq 1.25$.
At three loops we can use the result (\ref{b2oms}) to repeat the game
with the next-to-next-to--leading logarithm, which is multiplied by
$d_{31}=9736.8$ in (\ref{coup}): $|d_{33} \ln^3 (\mu_0^2/M_H^2 )|$ and
$|d_{32} \ln^2 (\mu_0^2/M_H^2 )|$ are smaller than $|d_{31} \ln
(\mu_0^2/M_H^2 ) |$ for $0.3 \leq \mu_0/M_H \leq 3.3$. For
$\mu_0/M_H=0.3$ or $\mu_0/M_H=3.3$ this term, however, exceeds $2\cdot
10^4$, which is above the value one expects for the yet unknown
constant $|d_{30}|$. Since $d_{33},d_{31} > 0$ and $d_{32} < 0$ the
choice $\mu_0/M_H =0.7 < 1 $, which has been so seductive in the order
$\hloms^1$, leads to the fact that the three logarithmic terms in the
order $\hloms^3$ add with the same sign to yield $-8045$. Moreover
from (\ref{llog}) and (\ref{nllog}) one realizes that for $\mu_0 <M_H$
the leading and next-to-leading logarithmic terms have the same sign
to all orders $\hloms^n$ except for $n=1$ and $n=2$!  Hence the
one-loop FAC choice $\mu_0 = 0.7 M_H$  pushes large terms into the
higher orders. The two-loop FAC scale is even lower, increasing
all higher order terms even more.

We conclude that our lack of knowledge of the higher order terms of
the perturbation series forces us to consider {\em any}\/ choice for
$\mu_0$ in the range $0.8 \leq \mu_0/M_H \leq 1.25$ (which may
possibly be relaxed to $0.5 \leq \mu_0/M_H \leq 2.0$) with equal
right. Changing $\mu_0$ in the $n$-th order perturbative expression
for some observable changes the result by terms of the neglected order
$\lambda^{n+1}$. When perturbation theory works the dependence on
$\mu_0$ diminishes order-by-order in $\lambda$.  We will use a
similar criterion to find the breakdown of perturbation theory in
sections \ref{sect:scale} and \ref{sect:cross}.

Let us close this section with a final remark on the arbitrariness
of $\mu_0$:
Cross sections with cm-energy $\sqrt{s}$ and expressed in terms of
$\lams (\mu)$
involve the logarithm $\ln (\mu^2/s)$. Using (\ref{coup})
with $\mu = \mu_0$, this logarithm would be large.  Using the
running coupling $\lams (\mu)$ evaluated at a scale $\mu\approx
\sqrt{s}$, the logarithm is summed to all orders in
perturbation theory. The arbitariness in the choice of $\mu_0\approx
M_H$ and $\mu \approx \sqrt{s}$ reflects the fact that one can sum an
arbitrary small constant together with the large logarithm. This
feature is also present in the OMS scheme, but less apparent. For
example, the authors of \cite{lend} sum the constant $c_1$ in
(\ref{coup}) together with $\ln (\mu^2/s)$.

\subsection{The NNLL $\beta$-function in the OMS and
$\overline{\rm MS}$-scheme }

Neglecting all couplings except for the Higgs coupling $\lambda$, the
Higgs sector of the Standard Model is equivalent to a spontaneously
broken $\phi^4$ theory with $N=4$ real scalar fields.  Theories with
spontaneous symmetry breaking have the remarkable property that the
counterterms needed to make the theory finite are the same for the
broken and unbroken symmetry, if a mass independent renormalization
scheme is adopted \cite{collins}.  Since the beta coefficients are
calculated from the counterterms of the coupling, Eq.~(\ref{z}), the
beta function is also identical for both the broken and the unbroken
theory.  Hence one can calculate the $(n+1)$-loop coefficient
$\beta_{n}^{\rm OMS}$ of the Higgs sector in two steps: First, obtain
$\beta_n^{\overline{\rm MS}}$ in an unbroken $\phi^4$ theory with
$N=4$ real scalar fields by calculating the divergent parts of the
four-point function to the $(n+1)$-loop order.  This is easier than
working in the broken theory, because only a four-point coupling is
involved.  Once this is accomplished, $\beta_n^{\overline {\rm MS}}$
of the broken theory is known, too. Second, calculate the scheme
dependence $\beta_n^{\overline{\rm MS}}-\beta_n^{\rm OMS}$. This only
requires the calculation of the $n$-loop (not $n+1$-loop) finite parts
of self-energy diagrams.

The three-loop coefficient $\beta_2$ of the $\phi^4$
theory has been calculated in the $\overline{\rm MS}$ scheme
\cite{vlad}:
\begin{eqnarray}
\beta_2^{\overline{\rm MS}}&=&
7176 + 4032\zeta(3)  \; = \;  12022.69\ldots \,.
\end{eqnarray}
To obtain $\beta_2^{\rm OMS}$, we return to Eq.~(\ref{z}) to calculate
the scheme dependence of the coefficient $\beta_2$.  As stated above,
unlike $\beta_2$ itself the scheme dependent difference between
$\beta_2^{\overline{\rm MS}}$ and $\beta_2^{\rm OMS}$ can entirely be
obtained from two-loop quantities:
\begin{eqnarray}
\beta_2^{\overline{\rm MS}}&=&\beta_2^{\rm OMS} - \beta_1 c_1
+\beta_0 c_2 - \beta_0 c_1^2 - \beta_0^2c_{11}\nonumber\\
&=&\beta_2^{\rm OMS}
- 5400 + 8688\zeta(2) - 2160\zeta(3) - 2736\pi\sqrt{3}
\nonumber \\
&& +1152\pi{\bf{Cl}}(\pi/3)
+5184\sqrt{3}{\bf{Cl}}(\pi/3) + 3888 K_5
\nonumber\\
&=&\beta_2^{\rm OMS} + 7784.45\dots   \label{schemeres}
\end{eqnarray}
where $K_5 = 0.92363\dots$ is the value of the all massive Master diagram
with $p^2=M_H^2$,
which has been evaluated numerically in \cite{maher}.
(\ref{schemeres})  can be obtained from the definition
(\ref{beta}) of the $\beta$-function or by comparing the bare coupling
expressed in terms of $\lams$ with its expression in terms of $\loms$.

Using the result for $\beta_2^{\overline{\rm MS}}$ we obtain
\begin{eqnarray}
\beta_2^{\rm OMS} &=&    4238.23\dots \label{b2oms} \,.
\end{eqnarray}
Our analytical\footnote{apart from the numerical constant $K_5$ which
is defined by a single Feynman diagram}
result agrees with the numerical result obtained in
\cite{luewei2} to better than six digits.

\begin{figure}[tbh]
\vspace*{13pt}
\centerline{
\epsfysize=4.8in \rotate[l]{\epsffile{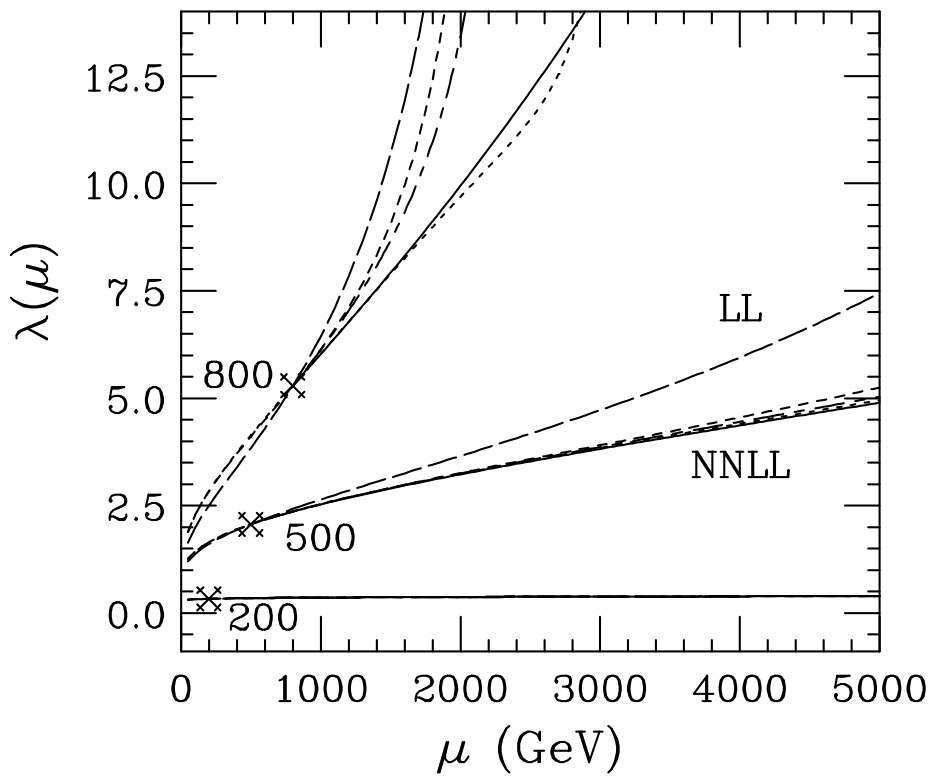}}
}
\vspace{0.15in}
\caption{
Different solutions of the three-loop RGE equations for the
Higgs quartic coupling $\lambda$:
the NNLL (three-loop) naive solution according to (\protect\ref{analint2})
(long-short dashes), the NNLL consistent solution of
(\protect\ref{consist2lp}) (solid line), the NNLL iterative solution
(short dashes), and the NNLL QCD-like solution
expanded in powers of $1/\ln(\Lambda_H^2/\mu^2)$ given in
(\protect\ref{lamdef}) (dots).
The symbols and labels are the same as in
Fig.~{\protect\ref{rgecurves}}.  For comparison,
the LL (one-loop) solution (long dashes) is shown again.
}
\label{rgecurves3lp}
\end{figure}

In Fig.~\ref{rgecurves3lp} we show the $\mu$-dependence of the NNLL
(three-loop) running coupling in the OMS scheme. The consistent
solution is almost identical to its NLL (two-loop) result even for
$M_H=800$ GeV, whereas the other NNLL solutions show a behaviour very
different from their NLL results for large $M_H$.  For $M_H=500$ GeV
and $\mu<5$ TeV all four NNLL solutions show a very nice
convergence. For such a value of $M_H$, however, the LL running
coupling is not an adequate approximation for the upper values of
$\mu$ considered.

In the ${\overline {\rm MS}}$ scheme, the convergence of the running
coupling is rather poor for $M_H=500$ GeV and above. This is also true
for the consistent solution.  For $M_H=750$ GeV, the consistent
NNLL solution of the ${\overline {\rm MS}}$ running coupling is
not defined anymore if $\mu>950$ GeV.  The poor performance of the
NNLL ${\overline {\rm MS}}$ coupling is due to the term involving
$\beta_2$. Although $\beta_2^{\overline {\rm MS}}$ is only three times
larger than $\beta_2^{\rm OMS}$ the coefficient
$\beta_1^2-\beta_0\beta_2$ entering the running coupling
(\ref{consist2lp}-\ref{lamdef}) is 44 times larger in the
${\overline {\rm MS}}$ scheme. This is caused by a numerical
cancellation in the OMS scheme, where
$\beta_2^{\rm OMS}/\beta_1 \approx \beta_1/\beta_0$. We remark here
that in a scheme with exact geometrical growth of the coefficients,
$\beta_{n+1}/\beta_n=\beta_1/\beta_0$, the consistent solution
(\ref{consist2lp}) equals the consistent NLL result to all orders.

\section{SCHEME AND SCALE DEPENDENCE OF HIGGS DECAYS}\label{sect:scale}
The accuracy of perturbation theory and its breakdown as $M_H$
increases can only be investigated in physical observables.  Here
processes in which all mass parameters are of the order of $M_H$ are
of key importance: If the Higgs self-interaction is non-perturbative
at the scale $M_H$ at which (\ref{defloms}) is imposed, we do not
expect perturbation theory to work in any other observable.  Two-body
Higgs decay rates are examples for such one-scale processes.  They do
not contain large logarithms, which need to be summed to all orders.
Nevertheless as emphasized at the end of Sect.~\ref{sect:mscoup}, RG
methods can be used to judge the accuracy of perturbation theory: The
dependence on the renormalization scale $\mu$, at which the decay rate
is evaluated, is of the neglected order in $\lambda$. The stability of
the perturbative result with respect to variations of $\mu$ must
therefore increase order-by-order in $\lambda$.  This criterion to
test perturbation theory has been first used in QCD in \cite{bjw} and
has become a standard method in QCD.  Likewise the ratio of two
results obtained in different renormalization schemes must approach unity
with increasing order in $\lambda$.  This renormalization scheme
dependence will be the second tool used in the analysis of the
breakdown of perturbation theory.

\subsection{Higgs decay into gauge bosons}
At first we consider the decay rate of a Higgs into two gauge bosons.
The ${\rm O}(\lambda^n)$ corrections to the decay rate $H\rightarrow W^+W^-$
are the same as the ones for the decay into two $Z$ bosons. The decay
rates of the two channels only differ by an overall factor of
proportionality.

We write for the  decay rate
\begin{eqnarray}
\Gamma (H\rightarrow W^+W^-) &=& \Gamma_{\rm tree}
    \frac{\lambda_{\rm scheme}}{\lambda_{\rm tree}} \left( 1 \, + \,
                                   \Delta\Gamma_{\rm scheme}
                \right) .  \label{defww}
\end{eqnarray}
Here $\Gamma_{\rm tree}$ is the Born approximation with
\begin{equation}
\Gamma_{\rm tree} = M_H
\sqrt{1-\frac{4M_W^2}{M_H^2}}
\left(1-\frac{4M_W^2}{M_H^2}+\frac{12M_W^4}{M_H^4}\right)\,
\frac{\lambda_{\rm tree}}{8 \pi} \, .
\end{equation}
The corrections stemming from the Higgs self-interaction have been
calculated in the OMS scheme to two loops~\cite{hww}:
\begin{eqnarray}
1\, +\, \Delta\Gamma_{\rm OMS}
 &=& 1\, +\, 2.80\hat{\lambda}_{\rm OMS}\, +\, 62.15\hat{\lambda}_{\rm
 OMS}^2\,  +\, {\rm O}\left(\hat{\lambda}_{\rm OMS}^3\right)
+\, {\rm O}\left(\frac{M_W^2}{M_H^2}\right) \, .
\label{hwwoms}
\end{eqnarray}
In \cite{hww} it has been stressed that the two-loop OMS correction
exceeds the one-loop OMS term for $M_H>930$ GeV, although the one-loop
correction is still small compared to the tree-level term. This
indicates that either perturbation theory does not work for $M_H>930$ GeV or
the one-loop term is accidentally small in the OMS scheme.

Using (\ref{coupinv}) we can express $\Gamma$ in terms of the
$\overline{\rm MS}$ coupling:
\begin{eqnarray}
1\, +\, \Delta\Gamma_{\overline{\rm MS}} &=& 1\, - \,
 \hat{\lambda}_{\overline{\rm MS}}\,
\left( 12 \ln \frac{\mu^2}{M_H^2}  + 5.88    \right)
+\, \hat{\lambda}_{\overline{\rm MS}}^2\,
\left( 144 \ln^2 \frac{\mu^2}{M_H^2}  + 297.0 \ln \frac{\mu^2}{M_H^2}
       -122.7 \right) \nonumber \\
&& +\, {\rm  O}
   \left(\hat{\lambda}_{\overline{\rm MS}}^3\right) +\, {\rm O}
   \left(\frac{M_W^2}{M_H^2}\right) \, . \label{hwwmsbar}
\end{eqnarray}
Here $\mu$ is the renormalization scale at which the decay rate is
evaluated. The renormalization point $\mu_0$, at which the coupling
is defined by (\ref{coup}) in terms of $G_F$ and $M_H$, is chosen
as $\mu_0=M_H$ throughout this section.
If we also take $\mu=M_H$, we observe that the
apparent convergence of $\Delta\Gamma_{\overline{\rm MS}}$ is worse
than that of $\Delta\Gamma_{\rm OMS}$ since both $\lams > \loms$ and
the ${\overline {\rm MS}}$ coefficients of the perturbation series are
in magnitude larger than in the OMS scheme.  The two-loop ${\overline
{\rm MS}}$
correction equals the one-loop ${\overline {\rm MS}}$ term
for $M_H=770$ GeV. The corresponding OMS result yields $M_H=930$ GeV.

Let us now investigate whether one can refine these bounds on a
perturbative Higgs mass by the examination of the renormalization
scheme and scale dependence.

\begin{figure}[tb]
\vspace*{30pt}
\centerline{
\epsfysize=3.0in \rotate[l]{\epsffile{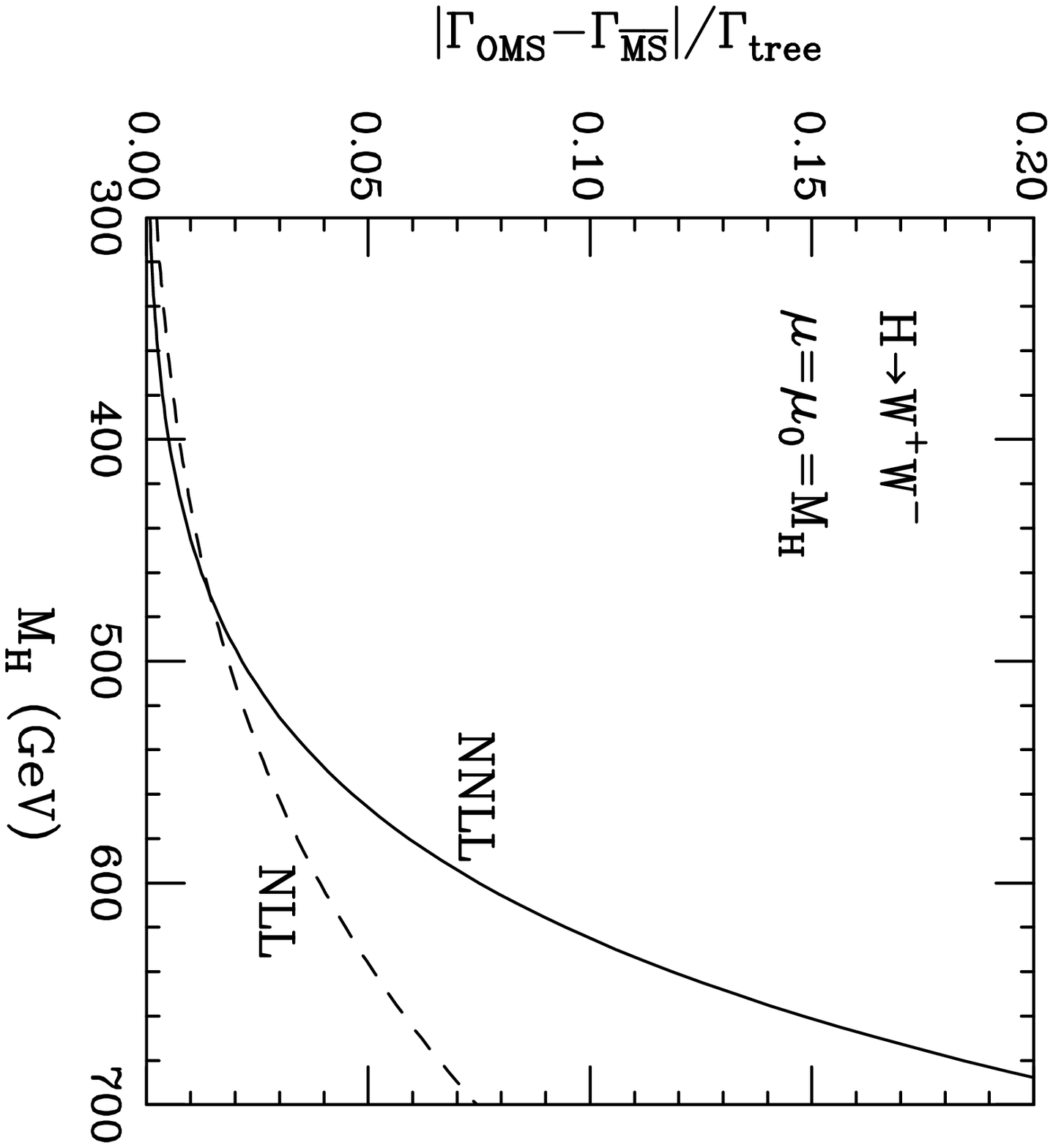}}
\epsfysize=3.0in \rotate[l]{\epsffile{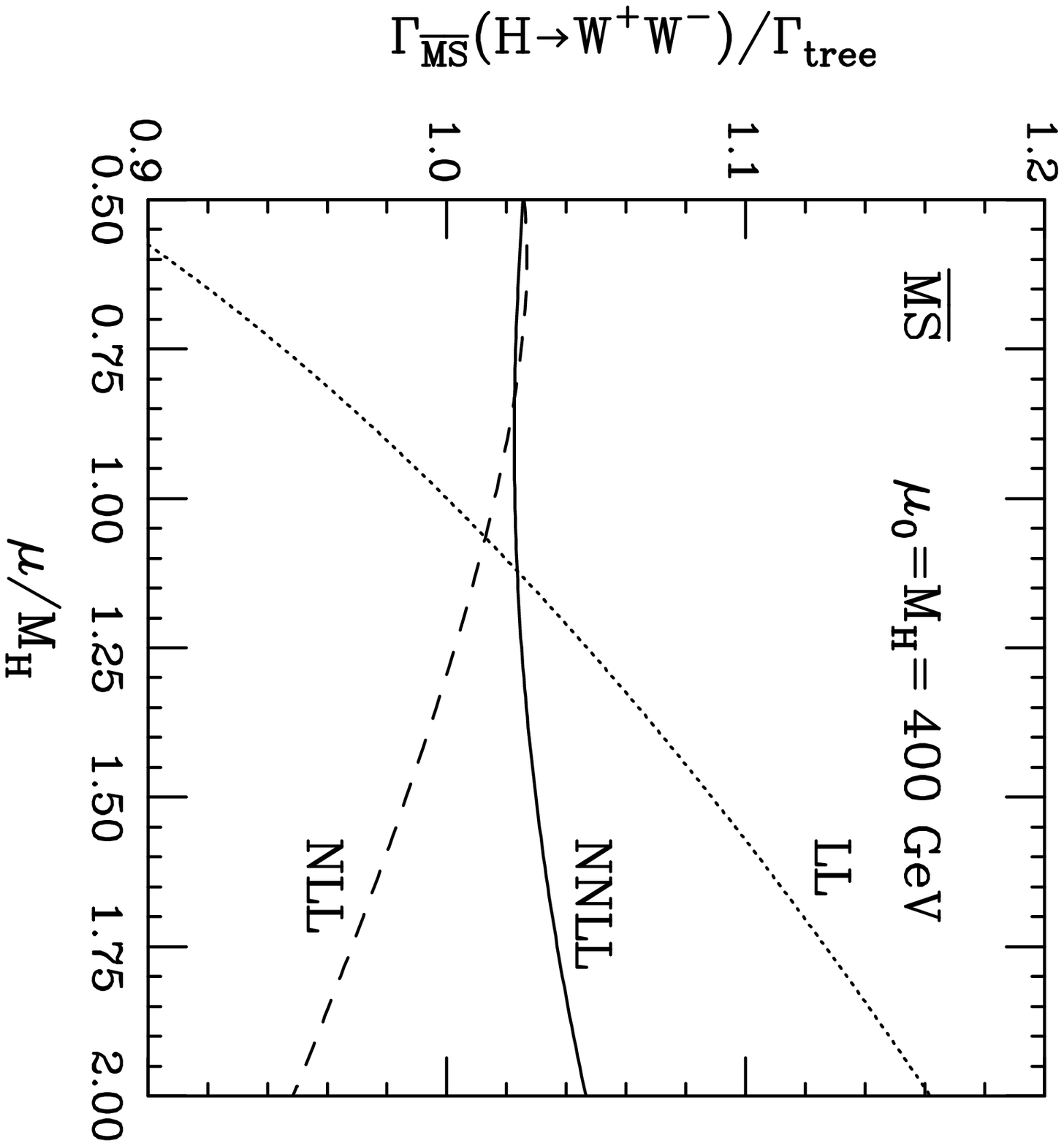}}
}
\vspace{0.15in}
\caption{The left plot shows the scheme dependence
$(\Gamma_{\text{OMS}}-\Gamma_{\overline{\text{MS}}})/\Gamma_{\rm
tree}$ of the decay rate
$\Gamma (H \rightarrow W^+ W^-)$. For $M_H > 469$~GeV the two--loop
scheme dependence (solid line) is larger than the one--loop scheme
dependence (dashed).  The right plot shows the dependence on the
renormalization scale $\mu$ of $\Gamma_{\overline{\text{MS}}}$ for $M_H=
400$ GeV. The scale dependence diminishes sizably order-by-order
indicating that perturbation theory works well for this value of $M_H$.}
\label{figscheme}
\end{figure}

We start our analysis by looking at the scheme dependence of the decay width.
In the left plot of figure Fig.~\ref{figscheme}
we show the normalized scheme dependence
$(\Gamma_{\rm OMS}-\Gamma_{\overline{\rm MS}})/\Gamma_{\rm tree}$ at
one and two loops, using NLL and NNLL running couplings, respectively.
The plot indicates that for $M_H < 469$ GeV the scheme dependence
is reduced when going from one-loop to two-loop
order.  For larger Higgs masses, the scheme dependence
increases, suggesting that perturbation theory is not working
satisfactory or even fails. The scheme-dependence criterion indicates
problems with perturbation theory in
at least one of the two schemes considered for $M_H > 469$ GeV.
The criterion, however, is sensitive to the possibility of an
accidental smallness of the one-loop correction.
We remark here that the two-loop ${\overline {\rm MS}}$ term in
(\ref{hwwmsbar}) is less than four times the square of the one-loop
term, while in the OMS scheme, (\ref{hwwoms}), this ratio almost
equals eight.  For $M_H=700$ GeV the result for $\Gamma_{\rm OMS}$ is
larger than the $\overline{\rm MS}$ expression by 23\%, and for
$M_H=780$ GeV the scheme dependence reaches unacceptable 52\%.  The
simple criterion of comparing the magnitudes of the one-loop and
two-loop contributions gives upper bounds on $M_H$ which seem to be
too large.

Another criterion for the validity of perturbation theory is the
order-by-order reduction of the scale dependence.  The explicit
$\mu$-dependent logarithms in (\ref{hwwmsbar}) compensate the effect of
the running coupling $\lambda(\mu)$ to the order considered.  To use
this criterion in the OMS scheme as well, we need to introduce the RG
logarithms into the OMS decay width (\ref{hwwoms}).  In the OMS
scheme, the Callan-Symanzik equation describes the response of some
Green's function to the scaling $p_j \rightarrow \mu/M_H \cdot p_j$ of
its external momenta. Its solution for the decay rate (\ref{hwwoms})
reads
\begin{eqnarray}
1\, +\, \Delta\Gamma_{\rm OMS} &=& 1\, +\,   \hat{\lambda}_{\rm OMS}
 (\mu) \left( -12 \ln \frac{\mu^2}{M_H^2} + 2.80  \right)
  \nonumber \\
&&  +\, \hat{\lambda}_{\rm OMS}^2 (\mu)
  \left( 144 \ln^2 \frac{\mu^2}{M_H^2}
+  88.8 \ln \frac{\mu^2}{M_H^2}+ 62.15  \right)  \nonumber\\
&&  +\, {\rm
 O}\left(\hat{\lambda}_{\rm OMS}^3\right) +\, {\rm
 O}\left(\frac{M_W^2}{M_H^2}\right) \, . \label{hwwcs}
\end{eqnarray}
Expanding the running coupling $\loms (\mu)$ in (\ref{defww}) and
(\ref{hwwcs}) in terms of $\loms (M_H) $, (\ref{hwwcs}) yields
(\ref{hwwoms}) up to the neglected order $\loms^3$, of course.  The
OMS result expressed in terms of the running coupling (\ref{hwwcs})
can be used to examine the scale dependence of the decay width.

We use the LL, NLL and NNLL expressions of the decay rates which
consist of the Born, one- and two-loop result in (\ref{hwwmsbar})
supplied with the LL, NLL or NNLL running coupling $\lams (\mu)$,
respectively.  Then the renormalization scale $\mu$ is varied.  The
right plot in Fig.~\ref{figscheme} shows that perturbation theory
nicely works for the $\overline{\rm MS}$ scheme if $M_H=400$ GeV: The
scale dependence diminishes order-by-order.  A similar behavior is
found for the OMS result using the OMS running coupling and the same
value of $M_H$.  Next we look for an upper bound for a perturbative
Higgs mass. We investigate the scale dependence of the decay rate for
values of $M_H$ up to 800 GeV.  Based on the size of the logarithmic
terms in the higher orders of the perturbation series for the $\lams$
we have already advocated the range $0.8 M_H < \mu < 1.25 M_H$ in
Sect.~\ref{sect:mscoup}. We may add a physical argument for this range
as well: Suppose one decides to include the non-zero width of the
Higgs into the analysis. Then the decay diagrams must be calculated
with an off-shell Higgs boson with invariant mass $\sqrt{s}$, and the result
is convoluted with a Breit-Wigner function. The decay rate would
differ from (\ref{hwwmsbar}) by a function of $s$ and $M_H$ vanishing
for $s=M_H^2$. The decay rate then involves two logarithms: $\ln
(\mu^2/s)$ and $\ln (\mu^2/M_H^2)$. The choice of $\mu$ could be with
equal right $\mu = M_H$, $\mu=\sqrt{s}$ or any scale in between.  This
suggests choosing the range for $\mu$ to be of the order of the total
width.  For the values of Higgs masses we are interested in, the width
is between $0.2 M_H$ and $0.3 M_H$
\cite{hww}, so that the range $0.8 M_H < \mu < 1.25 M_H$ is
appropriate in both OMS and ${\overline {\rm MS}}$ scheme.

\begin{figure}[tb]
\vspace*{30pt}
\centerline{
\epsfysize=3.0in \rotate[l]{\epsffile{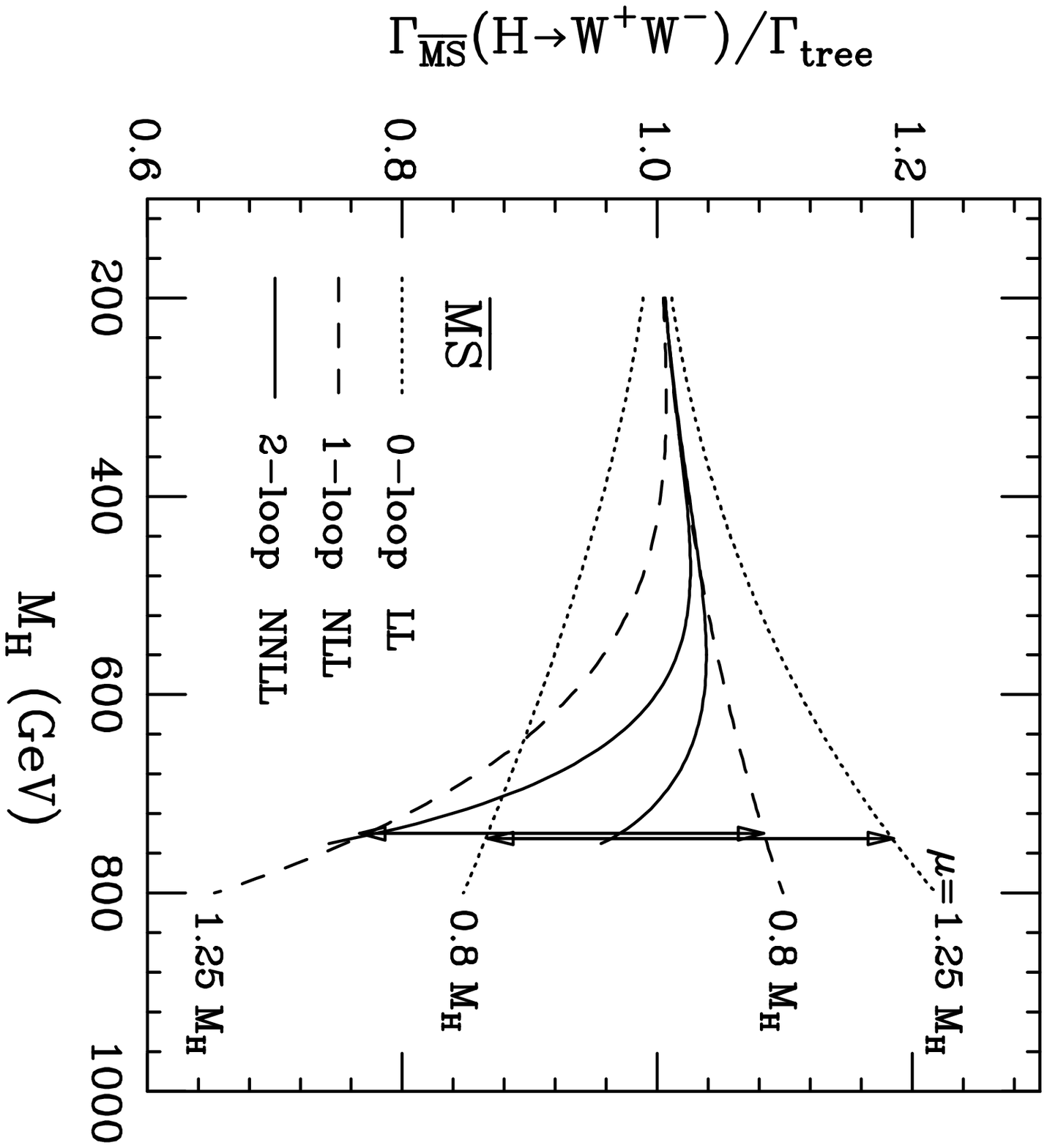}}
\epsfysize=3.0in \rotate[l]{\epsffile{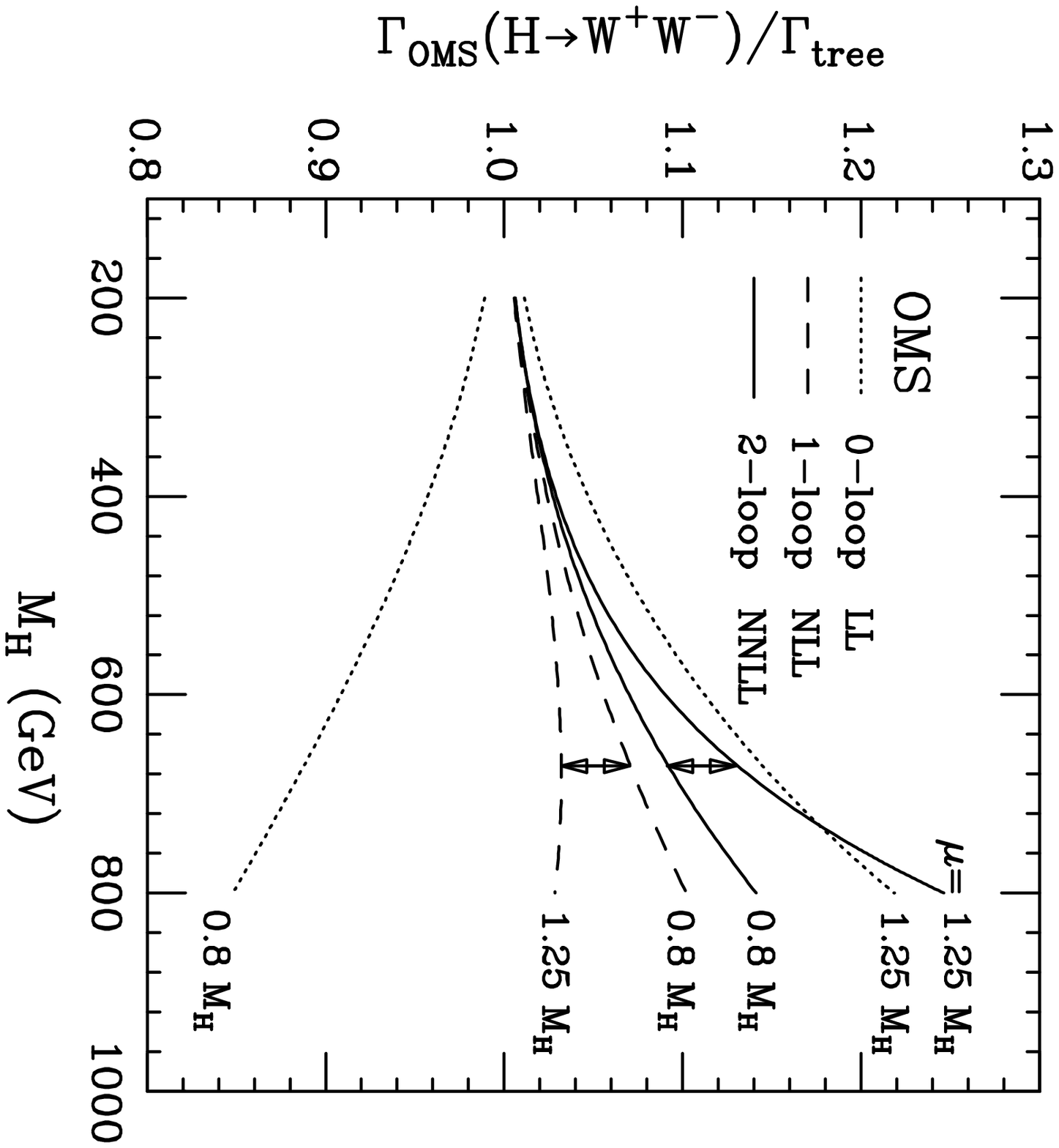}}
}
\vspace{0.15in}
\caption{The scale dependence of $\Gamma /\Gamma_{\rm
tree}$ for $H \rightarrow W^+ W^-$ in the $\overline{\text{MS}}$ (left
plot) and OMS (right plot) scheme. The extremal values of $\Gamma
/\Gamma_{\rm tree}$ are plotted for $0.8 M_H \leq \mu \leq 1.25 M_H$
in the LL (dotted), NLL (dashed) and NNLL (solid) approximation.
Except for the $\overline{\text{MS}}$ NNLL solution the extremal
values correspond to the chosen bounds $\mu=0.8 M_H$ and $\mu=1.25
M_H$ as indicated in the figures.  The scale dependence diminishes
order-by-order for $M_H<742$ GeV in the $\overline{\text{MS}}$ scheme
and for $M_H<672$ GeV in the OMS scheme. The arrows indicate the
equality of the LL and NLL scale uncertainty in the
$\overline{\text{MS}}$ scheme (left plot), and the equality of NLL and
NNLL scale uncertainty in the OMS scheme (right
plot). }\label{figscale}
\end{figure}

In Fig.~\ref{figscale} we have plotted the scale dependence of $\Gamma
(H\rightarrow W^+W^-)/(\Gamma_{\rm tree})$ vs.\ the
physical Higgs mass in both OMS and ${\overline {\rm MS}}$ scheme.
The tree-level coupling is $\lambda_{\rm tree}=G_F
M_H^2/\sqrt{2}$. The scale dependence at a given order is represented
by the smallest and the largest value of $\Gamma/\Gamma_{\rm tree}$
when $\mu$ is varied in the range $0.8 M_H \leq \mu \leq 1.25 M_H$.
In the $\overline{\rm MS}$ scheme the scale
dependence correctly decreases when passing from LL to  NLL
to NNLL order if $M_H < 742$ GeV.  For $M_H=742$ GeV the  scale
dependence in the LL  and NLL order become equal and reach 36\%.
For $M_H>750$ GeV and $\mu=1.25M_H$, the consistent solution
of the NNLL running coupling is no longer defined in the ${\overline
{\rm MS}}$ scheme, indicating the ultimate breakdown.
In the OMS scheme, the NLL and NNLL scale dependences  become
equal for $M_H=672$ GeV, but are numerically small (8\%).

We conclude that perturbation theory for bosonic Higgs decays breaks
down for Higgs masses of the order of $700$ GeV.  The scale-dependence
criterion yields similar upper bounds on $M_H$ in both schemes,
although the absolute scale dependence is much smaller in the OMS
scheme than in the $\overline{\rm MS}$ scheme. Using running coupling
solutions other than the consistent one (\ref{consist2lp}), we obtain
similar bounds.

It should be noted that our scale-dependence criterion is not only
sensitive to the coefficients of the different orders in $\Delta \Gamma$, but
also to the coefficients of the $\beta$-functions which also enter
the non-logarithmic terms of the uncalculated higher orders via
diagrams connected with counterterms in (\ref{z}).

\subsection{Higgs decay into two fermions}
Our next example is the fermionic decay width of the Higgs particle.
At Born level it reads
\begin{equation}
\Gamma _{\rm tree}\left(H\rightarrow f\bar f\,\right)=
{N_cm_f^2M_H\over8\pi v^2}\left(1-{{4{m_f^2}}\over {M_H^2}}\right)^{3/2}.
\label{eqborn}
\end{equation}
Here $N_c=1$ (3) for lepton (quark) flavors.
At tree level, this process is independent of the Higgs coupling
$\lambda$. Including radiative corrections, the fermionic decay rate
receives corrections in powers of $\lambda$. In the OMS scheme
they are~\cite{gh,hfflong}
\begin{eqnarray}
\label{hffoms}
1\, +\, \Delta\Gamma_{\rm OMS}\left(H\rightarrow f\bar f\,\right)
 &\approx& 1\, +\, 2.12\hat{\lambda}_{\rm OMS}\, -\, 32.66\hat{\lambda}_{\rm
 OMS}^2\,  +\, {\rm O}\left(\hat{\lambda}_{\rm OMS}^3\right).
\end{eqnarray}

Since the tree-level result of the fermionic Higgs decay,
Eq.~(\ref{eqborn}), is independent of the coupling $\lambda$, we only
need the one-loop relation between ${\lambda}_{\rm OMS}$ and
$\lambda_{\overline{\rm MS}}$ to calculate the ${\overline{\rm MS}}$
decay width up to two loops. Likewise, our scale variation
criterion only involves LL and NLL running coupling in connection with
the one-loop and two-loop results. As a result we can only
compare the scale dependence of two instead of three orders.
Since the LL one-loop result is identical in both schemes, the scheme
dependence can only be compared at the NLL two-loop level. No
bounds on $M_H$ can be derived from the scheme-dependence criterion
without knowing the three-loop corrections.
This distinguishes the decay
$\Gamma \left(H\rightarrow f\bar f\,\right)$ from the case
$\Gamma \left(H\rightarrow W^+W^- \right)$ discussed in the preceeding
section.

Combining Eq.~(\ref{coup}) with the
previous equation, we obtain the correction to the fermionic Higgs
decay in $\overline{\rm MS}$ quantities \cite{willey,hffproc}:
\begin{eqnarray}
\label{hffmsbar}
1 + \Delta\Gamma_{\overline{\rm MS}}\left(H\rightarrow f\bar f\,\right)
 &\approx& 1 + 2.12\hat{\lambda}_{\overline{\rm MS}}\,
-\, \left(51.03 + 25.41\ln({\mu^2}/{M_H^2})\right)
\hat{\lambda}_{\overline{\rm MS}}^2\,
+\, {\rm O}\left(\hat{\lambda}_{\overline{\rm MS}}^3\right).
\end{eqnarray}

We also give the scale dependence of the OMS result:
\begin{eqnarray}
1\, +\, \Delta\Gamma_{\rm OMS} &=& 1\, +\, 2.12\hat{\lambda}_{\rm OMS}
 (\mu)
  -\, \left(  32.66
  + 25.41 \ln \frac{\mu^2}{M_H^2}  \right)
  \hat{\lambda}_{\rm OMS}^2 (\mu)
 +\, {\rm
 O}\left(\hat{\lambda}_{\rm OMS}^3\right) \, . \label{hffcs}
\end{eqnarray}

\begin{figure}[tb]
\vspace*{13pt}
\centerline{
\epsfysize=3.0in \rotate[l]{\epsffile{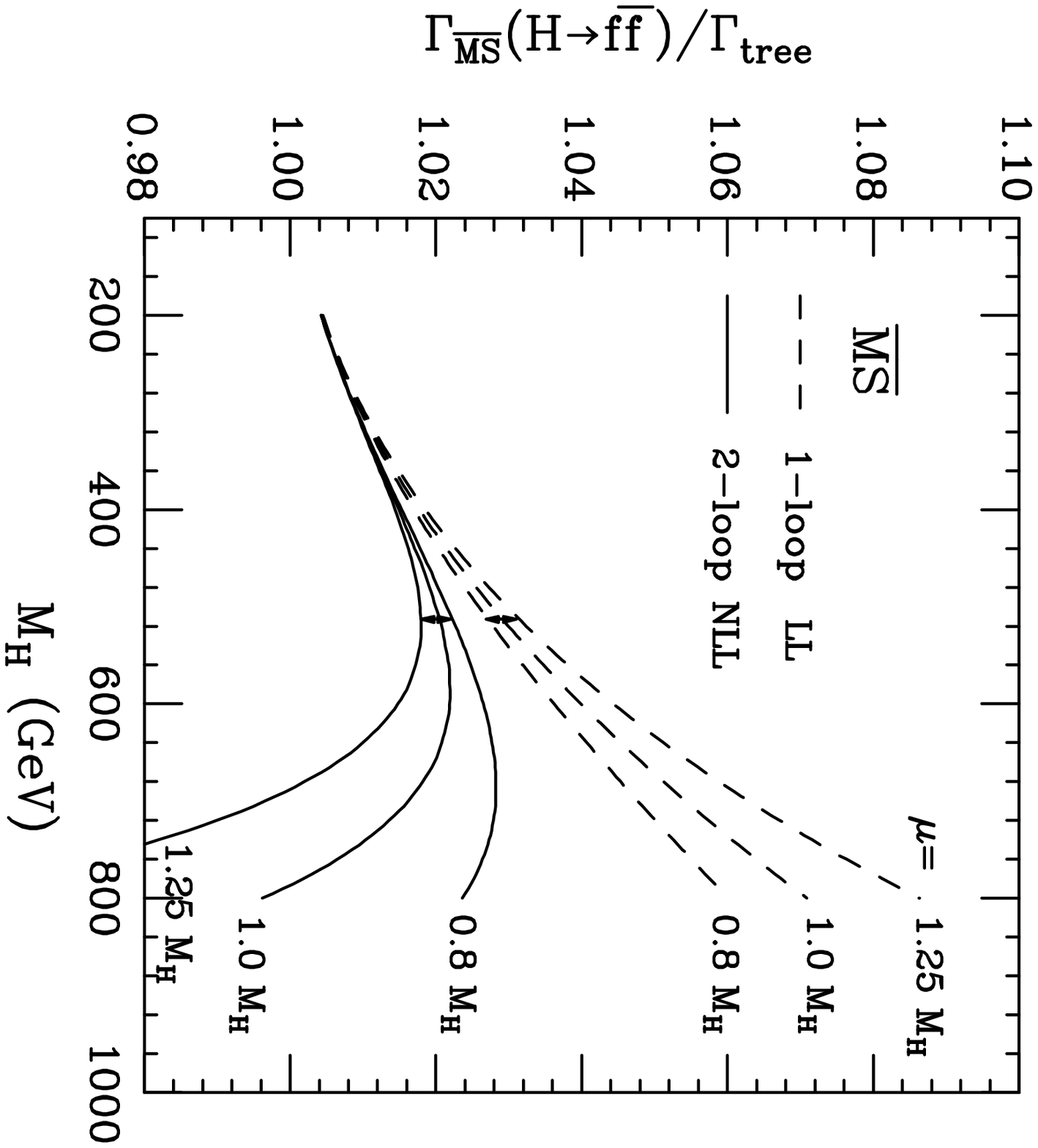}}
\epsfysize=3.0in \rotate[l]{\epsffile{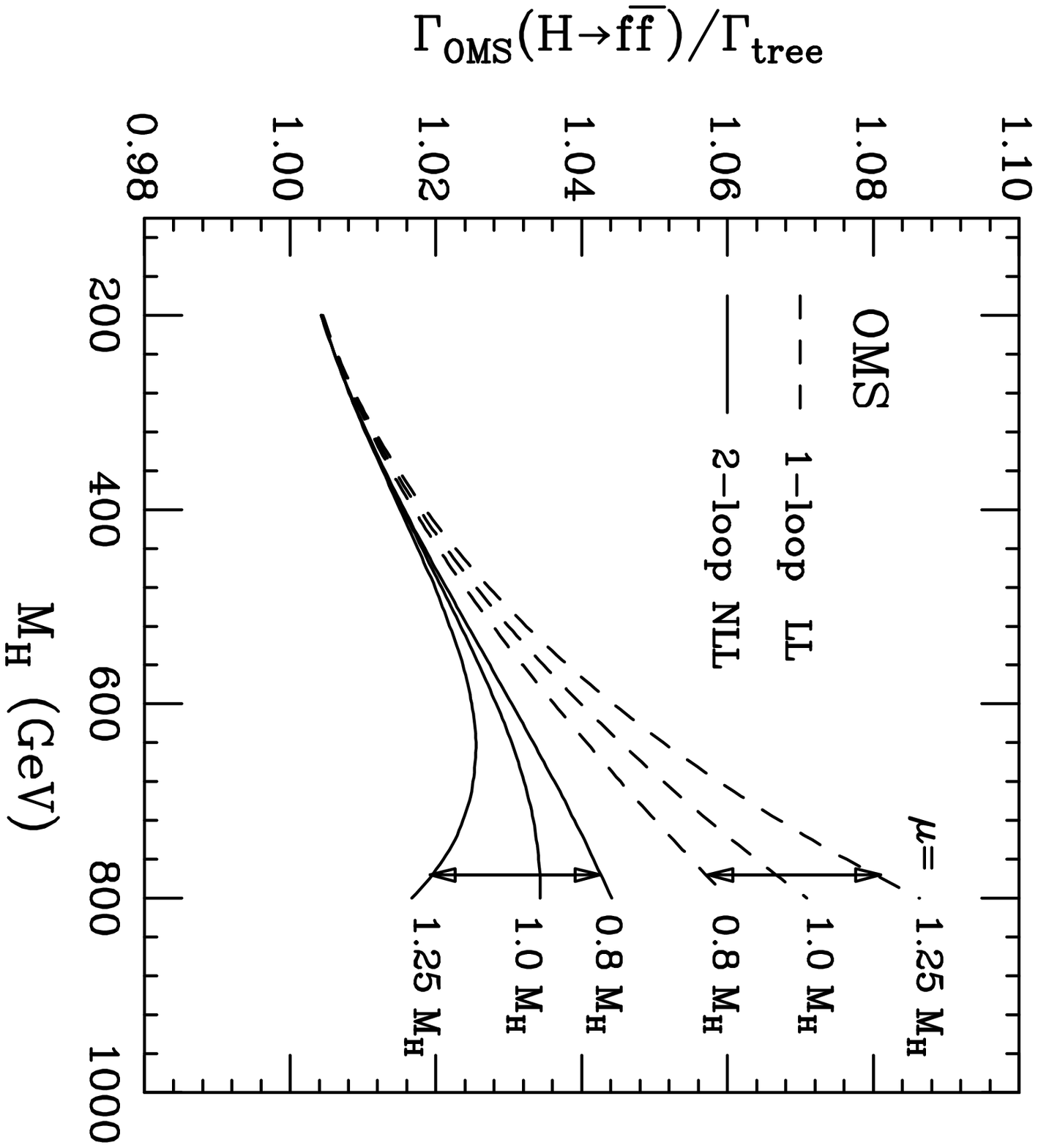}}
}
\vspace{0.15in}
\caption{
The resummed $\mu$-dependence of the normalized fermionic Higgs decay
width. Only the $\mu$-dependence entering through the Higgs coupling
is analyzed, i.e. no running Yukawa or QCD couplings are considered.
Using the $\overline{\rm MS}$ renormalization scheme (left plot), the
$\mu$-dependence of the 2-loop result is as big as the $\mu$-dependence of the
1-loop result for  $M_H=513$ GeV as indicated by arrows. In the OMS,
this equality occurs for $M_H=776$ GeV (right plot). The scale
dependence is nevertheless small at these critical Higgs mass values.
}
\label{hffrge}
\end{figure}

In Fig.~\ref{hffrge} we show the scale dependence of the decay width
expressed in terms of the running coupling in both ${\overline{\rm
MS}}$ and OMS scheme at one and two loops as a function of $M_H$. At
each order the three curves refer to $\mu= 0.8 M_H$, $1.0 M_H$, and
$1.25 M_H$.  We find that the $\mu$ dependence of the two-loop result
is larger than the $\mu$ dependence of the one-loop result if $M_H$ is
larger than $513$ GeV in the ${\overline {\rm MS}}$ scheme.  In the
OMS scheme the corresponding bound is $M_H=776$ GeV.  These results,
however, are not as valuable as those found in the previous section,
since they are only founded on the comparsion of two orders rather
than three. The low value of $M_H=513$ GeV in the ${\overline {\rm
MS}}$ scheme may be accidental.  The scale dependence is very weak,
less than a few percent for Higgs masses up to 750 GeV. This is due to
the fact that the tree-level result does not depend on $\lambda$.  The
scheme dependence at NLL (two-loop) is marginal for the same
reason. We presently have no information on the NNLL behaviour of the
fermionic decay.  Looking at Fig.~\ref{hffrge} we conclude that the
upper bound on a perturbative Higgs mass is in agreement with our
findings in the case of the bosonic Higgs decay.

\section{SCHEME AND SCALE DEPENDENCE OF CROSS SECTIONS}\label{sect:cross}
The previous section considered one-scale processes. The only
logarithms appearing in the RG improved results are
$\ln(\mu^2/M_H^2).$ Testing the perturbativity of Higgs sector is much
more stringent when considering high-energy two-scale processes,
because they involve the running coupling at a high scale.  Typical
two-scale processes which depend on the coupling $\lambda$ are
$2\rightarrow 2$ scattering processes involving longitudinally
polarized gauge bosons and the Higgs boson.  In the limit $s,M_H^2 \gg
M_W^2$ the longitudinally polarized gauge bosons can be replaced by
massless Goldstone bosons. In this limit all couplings except for the
trilinear and quartic Higgs coupling are subleading and can be
neglected.  Also assuming $s\gg M_H^2$, the high-energy amplitudes and
cross sections have been calculated to two loops
by~\cite{maher,kurt}. The high-energy limit is approached if
$\sqrt{s}>2 M_H$~\cite{kurt}. Its error is less than a few percent if
$\sqrt{s}>5 M_H$. For most of our analysis we will consider
$\sqrt{s}=2$ TeV.

A typical example is the process $W_L^+W_L^-\rightarrow Z_LZ_L$.
The high-energy cross section only depends on $\sqrt{s}$, $M_H$, and
$\lambda$. The OMS cross section is~\cite{kurt}
\begin{eqnarray}
\sigma(s)\,
& = & \frac{1}{8\pi s} [\lambda_{\rm OMS}(\mu)]^2\,
\Biggl[\,1\,+
\left( 24 \ln \frac{s}{\mu^2} -
       \, 42.65 \right) \,\frac{\lambda_{\rm OMS}(\mu)}{16\pi^2}\,
\nonumber \\
&& \phantom{\frac{1}{8\pi s} [\lambda_{\rm OMS}(\mu)]^2\, }
+ \left( 432 \ln^2 \frac{s}{\mu^2} - 1823.3\ln \frac{s}{\mu^2}
   + 24.0\ln \frac{s}{M_H^2}
   +\;2455.1\,\right) \frac{[\lambda_{\rm OMS}(\mu)]^2}{(16\pi^2)^2}\,
 \nonumber \\
&& \phantom{\frac{1}{8\pi s} [\lambda_{\rm OMS}(\mu)]^2\, }
+\; {\rm O}\left([\lambda_{\rm OMS}(\mu)]^3\right)\,\Biggr]\,,
\label{wwzzcross}
\end{eqnarray}
where the OMS renormalization fixes $\mu_0=M_H$ such that the OMS
running coupling has the boundary condition $\lambda_{\rm
OMS}(\mu=\mu_0)=M_H^2/(2v^2)$. The scale $\mu$ must be chosen of the
order $\sqrt{s}$ to resum the large logarithms to all orders. For a
complete resummation of all logarithms up to three loops, one uses the
NNLL running coupling and takes $\mu=\sqrt{s}$. The $\mu$-independent
logarithm is related to the field anomalous dimension and needs no
resummation because of the smallness of its coefficient. At one loop,
the anomalous dimensions of the Higgs and gauge bosons are zero, and
at two loops they are numerically unimportant~\cite{maher}.

The ${\overline {\rm MS}}$ result is
\begin{eqnarray}
\sigma(s)\,
& = & \frac{1}{8\pi s} [\lambda_{\overline {\rm MS}}(\mu)]^2\,
\Biggl[\,1\,+
\left(  24\ln\frac{s}{\mu^2} -\, 60.0 \right)\,
   \frac{\lambda_{\overline {\rm MS}}(\mu)}{16\pi^2}\,
\nonumber \\
&& \phantom{\frac{1}{8\pi s} [\lambda_{\overline {\rm MS}}(\mu)]^2\,}
+ \left( 432 \ln^2 \frac{s}{\mu^2} - 2472.0\ln \frac{s}{\mu^2} +
         24.0 \ln \frac{s}{M_H^2}
   +\;3367.9 \right) \,\frac{[\lambda_{\overline {\rm
   MS}}(\mu)]^2}{(16\pi^2)^2}\,
\nonumber \\
&& \phantom{\frac{1}{8\pi s} [\lambda_{\overline {\rm MS}}(\mu)]^2\,}
+\; {\rm O}\left([\lambda_{\overline
   {\rm MS}}(\mu)]^3\right)\,\Biggr]\,.
\label{wwzzmsbar}
\end{eqnarray}
We choose the renormalization point to be $\mu_0=M_H$ such that the
the ${\overline {\rm MS}}$ running coupling is fixed at $\mu=\mu_0$ by
(\ref{coup}).

The RG structure of the cross section is similar to the one of the
decay width $\Gamma(H\rightarrow W^+W^-)$ since the tree-level result
also depends on $\lambda$. There are two important differences: the
tree-level cross section is proportional to $\lambda$ {\it squared},
and the running coupling resums terms of order $\ln(s/M_H^2)$ which
can lead to a significant increase of the running coupling compared to
the tree-level coupling.

\subsection{Perturbativity at collider energies}
Choosing $\mu=\sqrt{s}=2000$ GeV, we show the scheme dependence of the cross
section $W_L^+W_L^-\rightarrow Z_LZ_L$ in Fig.~\ref{wwzzfigscheme}. For
Higgs masses larger than 400 GeV, the scheme dependence is larger than
40\% and actually increases when going from NLL to NNLL cross section.
Choosing $\mu=\sqrt{s}=1000$ GeV, a value more realistic for
$WW$-scattering at future colliders, we find the critical mass value
to be $M_H=436$ GeV. For such values of $M_H$ and $\sqrt{s}$ the
high-energy approximation yields about 70\% of the exact
$\lambda$-dependence of the cross section~\cite{kurt}.
Increasing $\sqrt{s}$, the breakdown of perturbation theory as seen in
the scheme-dependence criterion happens for smaller and smaller values
of $M_H$. We will come back to this later.

In addition to the scheme dependence, we also evaluate the
scale-dependence criterion.  Since the renormalization group is used
to resum $\ln(s)$ terms, the scale $\mu$ is varied around $\sqrt{s}$
rather than $M_H$.  The arguments of Sect.~\ref{sect:scheme}, which
are based on the size of the beta function coefficients, suggest the
range $0.8\sqrt{s}<\mu<1.25\sqrt{s}$.  Choosing $\sqrt{s}=2000$ GeV,
the result of this variation is shown in Fig.~\ref{wwzzrge} for both
${\overline {\rm MS}}$ and OMS scheme.  In the ${\overline {\rm MS}}$
scheme, we observe a nice order-by-order reduction of the scale
dependence if $M_H$ is less than 357 GeV. For larger values of $M_H$,
the NLL scale dependence exceeds the LL one, and for $M_H> 368$ GeV we
also find the NNLL scale dependence to be larger than the NLL one.
For the OMS result we find only one crossing point.  There the NNLL
scale dependence dominates over the NLL one if $M_H>410$ GeV.  Taking
also the scheme dependence into account we conclude that perturbative
results for longitudinal gauge boson scattering at $\sqrt{s}=2000$ GeV
cease to be meaningful if $M_H$ is of the order 400 GeV or above. For
a value of $\sqrt{s}=1000$ GeV we find limits of ${\rm O}(450\,
\text{GeV})$, but here the additional low-energy contributions ---
though not dominant --- may change the limit.

The upper bounds on a perturbative Higgs mass from scattering
processes are more stringent than the results found in Higgs decays.
The reasons are two-fold. First, the one-loop and two-loop
coefficients of the perturbative corrections to the cross section  are
an order of magnitude larger than the corresponding coefficients of
the decay processes. Second, the running coupling evaluated at a scale
$\sqrt{s}>M_H$ is numerically larger than the coupling involved in the
decay widths.

It is also interesting to note that the bounds derived here are similar
to the results found in \cite{unitar} which are based on
perturbative violation of unitarity at two loops, a rather different
criterion for the breakdown of perturbation theory.
Compared to the nonperturbative lattice results of $M_H<710\pm60$ GeV
\cite{heller}, our perturbative criteria require smaller Higgs masses.

\subsection{Perturbativity at large embedding scales}
\begin{figure}[tb]
\vspace*{30pt}
\centerline{
\epsfysize=4.0in \rotate[l]{\epsffile{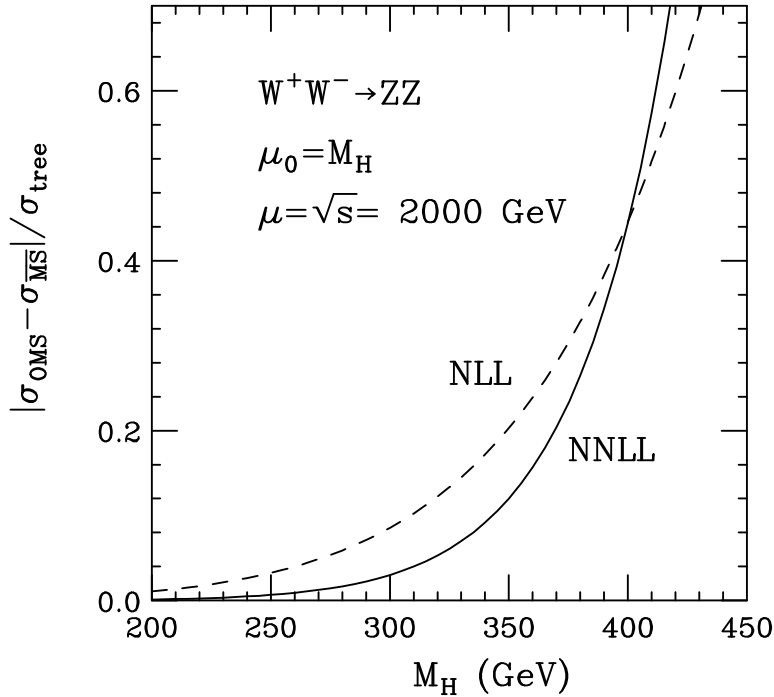}}
}
\vspace{0.15in}
\caption{The scheme dependence of the cross section
$W_L^+W_L^-\rightarrow Z_LZ_L$ at $\protect\sqrt{s}=2$ TeV.
For $M_H > 400$~GeV the two--loop
scheme dependence (solid line) is larger than the one--loop scheme
dependence (dashed) and exceeds 40\%.}
\label{wwzzfigscheme}
\end{figure}

\begin{figure}[tb]
\vspace*{13pt}
\centerline{
\epsfysize=3.0in \rotate[l]{\epsffile{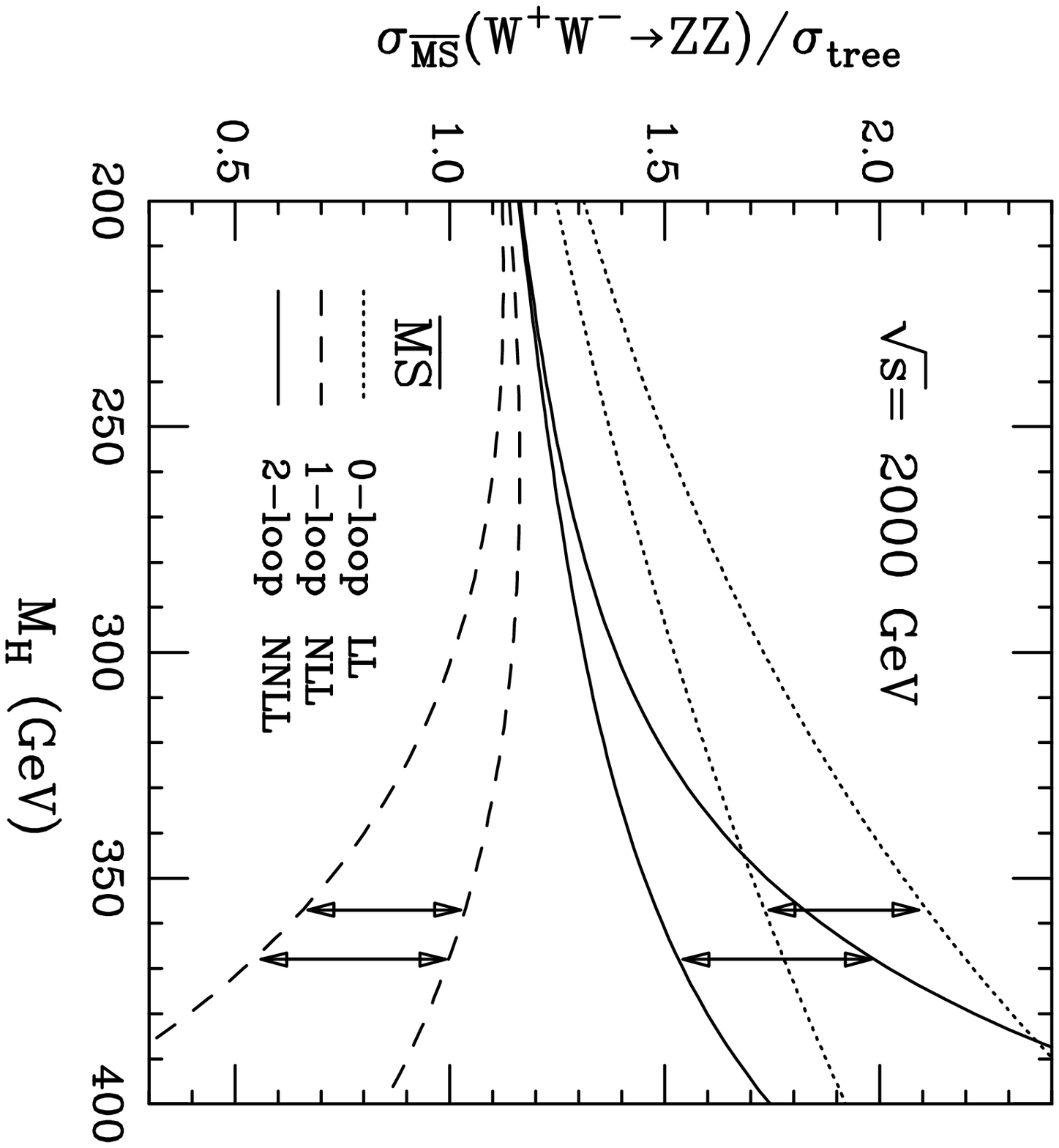}}
\epsfysize=3.0in \rotate[l]{\epsffile{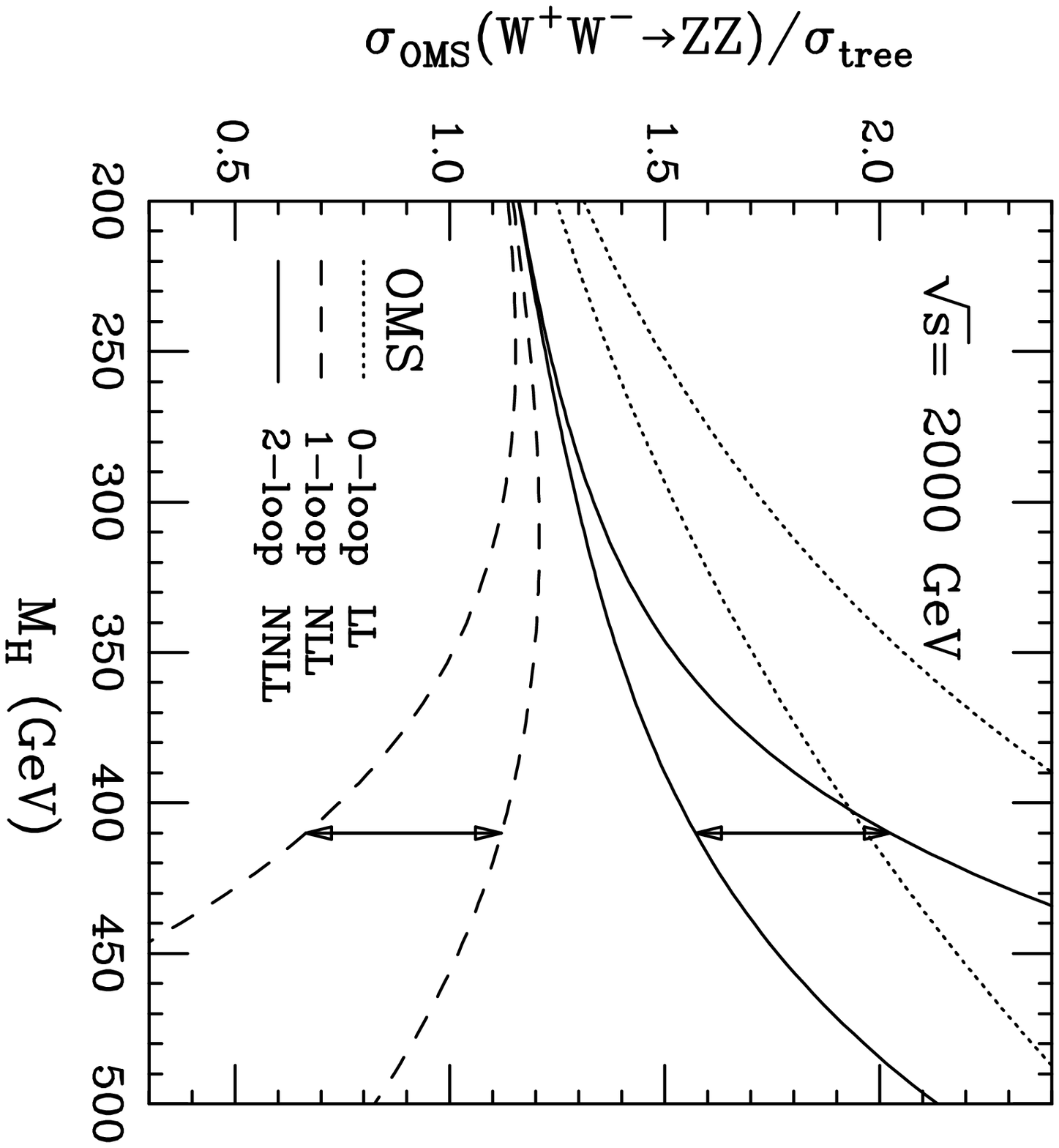}}
}
\vspace{0.15in}
\caption{
The resummed $\mu$-dependence of the normalized
$W_L^+W_L^-\rightarrow Z_LZ_L$ cross section at $\protect\sqrt{s}=2$ TeV.
Only the $\mu$-dependence entering through the Higgs coupling
is analyzed, i.e. no running Yukawa or QCD couplings are considered.
Using the $\overline{\rm MS}$ renormalization scheme,
only a Higgs mass of less than 357 GeV guarantees an order-by-order
reduction of the scale dependence. In the OMS scheme, the NLL and NNLL
are equal if $M_H=410$ GeV. The scale dependence is almost 50\% at
this value of $M_H$.
}
\label{wwzzrge}
\end{figure}

\begin{figure}[tb]
\vspace*{13pt}
\centerline{
\epsfysize=3.0in \rotate[l]{\epsffile{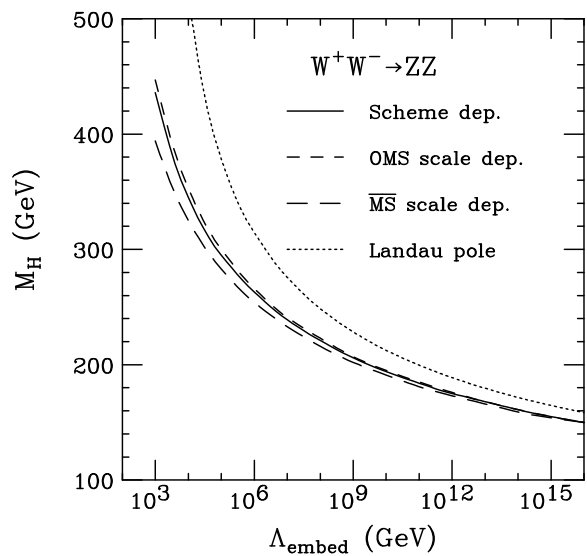}}
}
\vspace{0.15in}
\caption{
Upper bound on $M_H$ derived using our scheme-dependence and
scale-dependence criteria. We require the SM process
$W_L^+W_L^-\rightarrow Z_LZ_L$ to remain perturbative for cms-energies
up to $\protect\sqrt{s}=\Lambda_{\rm embed}$.
For comparison we also show the values of $M_H$ which lead to a
Landau singularity in the one-loop running coupling at the scale
$\mu=\Lambda_{\rm embed}$.
}
\label{wwzzembed}
\end{figure}

In Fig.~\ref{wwzzembed} we show the upper bound for a perturbative
Higgs mass as a function of some embedding scale $\Lambda$. Since the
Standard Model Higgs sector is --- depending on the value of $M_H$ ---
not defined above a certain energy scale, it is a common procedure to
introduce an embedding scale above which the physical interactions are
to be described by a more complete theory. Requiring that the SM Higgs
sector is still perturbative at such an embedding scale, it is
possible to give upper bounds on the Higgs mass as a function of the
embedding scale.  Here we require the process $W_L^+W_L^-\rightarrow
Z_LZ_L$ to be perturbative for energies $\sqrt{s}\leq\Lambda_{\rm
embed}$ and apply our criteria for scheme and scale dependence to
calculate the upper bound on $M_H$. The result is shown in
Fig.~\ref{wwzzembed}. At 2 TeV, the different bounds on $M_H$
correspond to the values derived from Figs.~\ref{wwzzfigscheme} and
\ref{wwzzrge}. For increasing embedding scale, the upper bound on
$M_H$ decreases, and the three different criteria give converging
bounds. At $\Lambda_{\rm embed}=10^{16}$ GeV, the upper bound is $150\pm
3$ GeV. For such low Higgs masses, however, one expects the top-quark
Yukawa coupling to have an influence on the RG evolution of
the Higgs coupling.  As a matter of fact, the SM beta function of the
Higgs coupling can become negative if the Higgs particle is too light,
invalidating our analysis.  Taking $m_t=175$ GeV, this is expected to
happen for $M_H<135$ GeV
\cite{altiso}, which is lower than the values considered by us.
Using the results of Lindner~\cite{lin}, we expect the change of our
Higgs mass bounds due to a top-quark mass of 175 GeV to be less than
10\% for the whole range of energy-scales considered.  For comparison,
we also display the values of $M_H$ which would lead to a one-loop
Landau singularity at $\Lambda_{\rm embed}$.  Calculating the Landau
pole, we again neglected Yukawa and gauge couplings.

It is worth noting that our criteria for an upper Higgs mass lead to
values of the running coupling at the embedding scale which are not
very large. For values of $\sqrt{s}$ in the TeV range, the maximal
allowed value of the running coupling is less than 2.2, and at GUT
energies we find the maximal value to be less than 1.6.  These small
values are surprising: The beta function and the solutions for the
running coupling show excellent convergence for such small values. As
a matter of fact, the usual criteria for the breakdown of perturbation
theory assume that the running coupling becomes large. Yet this method
is unsatisfactory: The consideration of the one-loop Landau pole as in
\cite{lin} cannot be extended to higher orders.
Instead the authors of \cite{lindner} use the criterion of (in our
notation) $\lambda(\Lambda_{\rm embed})<\pi^2$. Here the choice of the
numerical bound involves some arbitrariness. The plots in
Fig.~\ref{rgecurves} and Fig.~\ref{rgecurves3lp} show that $\pi^2$ is
clearly chosen too large, because the different solutions differ
substantially for $\lambda=\pi^2$. The second arbitrariness of this
methods resides in the fact that the breakdown of perturbation theory
is judged from a scheme dependent quantity: $\lambda(\Lambda_{\rm
embed})$ depends on the renormalization scheme through the radiative
corrections to the matching condition (\ref{defloms}) and, most
important, through the coefficients $\beta_n$, $n\geq 2$, of the
$\beta$-function. In contrast physical observables are scheme
independent up to the calculated order.  They, however, seem to become
non-perturbative for much smaller values of $\lambda$.  For comparison,
the unitarity arguments used in \cite{unitar} yield an upper bound on
the running coupling of 2.3, independent of any choice of $\sqrt{s}$.

\section{CONCLUSIONS}

Our analysis of various physical processes of the Higgs sector
indicates that the breakdown of perturbation theory cannot be judged
purely on grounds of a large Higgs (running) coupling. The breakdown of
perturbation theory in both Higgs decays ($\lambda_{\rm
OMS}=G_FM_H^2/\sqrt{2}<4$) and scattering
processes ($\lambda(\sqrt{s})<2.2$) occurs for relatively small values of the
Higgs coupling. The usual criterion --- the breakdown of the
perturbative behaviour of the beta function and the running coupling
--- yields upper bounds for a perturbative Higgs mass which are too
large.

Applying our criteria of scheme-dependence and
scale-dependence to Higgs decays, we find a satisfactory perturbative
behaviour of the decay widths if $M_H<$ O(700 GeV).
In the case of $2\rightarrow 2$ scattering processes, one needs to
specify the energy scale at which the process is to take place.
Choosing $\sqrt{s}$ to be a couple of TeV, the Higgs mass has to be less
than O(400 GeV) to guarantee a good perturbative behaviour of the
cross section for $W_L^+W_L^-\rightarrow Z_LZ_L$ scattering. Requiring
such processes to be perturbative at $\sqrt{s}=10^{16}$ GeV, the Higgs
mass has to be less than O(150 GeV).

\section{ACKNOWLEDGMENTS}
One of the authors (K.R.) gratefully acknowledges the hospitality of
the Aspen Center for Physics where parts of this work were done,
and thanks M. Lindner and S. Willenbrock for interesting and
informative discussions.


\begin{references}
\bibitem{jack} I.~Jack and H.~Osborn, J.\ Phys.\ {\bf A16}, 1101 (1983).
\bibitem{lend}
 E. Lendvai, G. P\'ocsik, and T. Torma, Mod.\ Phys.\ Lett.\ A
    {\bf  6}, 1195 (1991); Acta Phys.\ Pol. B {\bf 22}, 607 (1991).
\bibitem{bbdm}  W.~A.~Bardeen, A.~J.~Buras, D.~W.~Duke and T.~Muta,
               \pr D{\bf 18}, 3998 (1978).
\bibitem{landau}
L.D. Landau, in : Niels Bohr and the development of physics
(McGraw-Hill, New York, 1955).
\bibitem{petron}
L. Maiani, G. Parisi, and R. Petronzio, Nucl.\ Phys.\ {\bf B136}, 115 (1979);
\\ N. Cabbibo, L. Maiani, G. Parisi, and R. Petronzio, Nucl.\ Phys.\ {\bf
B158}, 295 (1979).
\bibitem{lindner}
B. Grzadkowski and M. Lindner, Phys.\ Lett.\ B{\bf 178}, 81 (1986).
\bibitem{luewei}
In the context of lattice studies this has been discussed in:
M. L\"uscher and P. Weisz, Phys.\ Lett. {\bf B212}, 472 (1988).
\bibitem{maher}
P.N. Maher, L. Durand, and K. Riesselmann,
   Phys.\ Rev.\ D {\bf 48}, 1061 (1993); (E) {\bf 52}, 553 (1995); \\
A. Ghinculov and J. van der Bij, Nucl.\ Phys.\ {\bf B436}, 30 (1995).
\bibitem{gh}
A. Ghinculov, Phys.\ Lett.\ B\ {\bf 337}, 137 (1994); (E) {\bf 346}, 426
   (1994).
\bibitem{sirzuc}
A. Sirlin and R. Zucchini, Nucl.\ Phys.\ {\bf B266}, 389 (1986).
\bibitem{steve} P.~Stevenson, Phys.~Rev.\ D{\bf 23}, 2916 (1981);
                Nucl.~Phys.\ {\bf B203}, 472 (1982).
\bibitem{collins}
J.C. Collins, {\it Renormalization} (Cambridge University Press, Cambridge,
1984), Chap.~9.
\bibitem{vlad}
A.A. Vladimirov, D.I. Kazakov, and O.V. Tarasov, Sov.~Phys.~JETP {\bf
50}, 521 (1979).
\bibitem{luewei2}
M. L\"uscher and P. Weisz, Nucl.~Phys.~{\bf B318}, 705 (1989).
\bibitem{bjw}  A.\ J.\ Buras, M.\ Jamin and
              P.\ H.\ Weisz, Nucl.\ Phys.\ {\bf B347}, 491 (1990).
\bibitem{hww}
A. Ghinculov, Univ.\ Freiburg preprint THEP 95/11, hep-ph/9507240.
\bibitem{hfflong}
L. Durand, B.A. Kniehl, and K. Riesselmann, Phys.\ Rev.\ Lett.\
{\bf 72}, 2534 (1994); (E) {\bf 74}, 1699 (1995);
Phys.\ Rev.\ D {\bf 51}, 5007 (1995).
\bibitem{willey}
A.I. Bochkarev and R.S. Willey,
   Phys.\ Rev.\ D {\bf 51}, 2049 (1995); Eq.~(17) of this paper uses incorrect
   numbers.
\bibitem{hffproc}
K. Riesselmann,
   in: Proceedings of the Ringberg Workshop on
   ``Perspectives for electroweak interactions in $e^+e^-$
   collisions'', Munich,
   Germany (Feb. 5-8, 1995), ed.:~B.~Kniehl,
   (World Scientific, Singapore, 1995), p.~175.
\bibitem{kurt}
K. Riesselmann, Tech.~Univ.~Munich preprint TUM-HEP-223/95 (July
1995), hep-ph/9507413.
\bibitem{unitar}
L. Durand, P. Maher, and K. Riesselmann, Phys.\ Rev.\ D{\bf 48}, 1084 (1993).
\bibitem{heller}
U. Heller, M. Klomfass, H. Neuberger, and P. Vranas, Nucl.\ Phys.\
{\bf B405}, 555 (1993).
\bibitem{altiso}
G. Altarelli and G. Isidori, Phys. Lett. {\bf B337}, 141 (1994).
\bibitem{lin}
M. Lindner, Z. Phys. {\bf C31}, 295 (1986).

\end{references}
\end{document}